\documentclass[twoside,11pt]{article}

 \usepackage{filecontents}
 
\usepackage{jmlr2e}

\usepackage[utf8]{inputenc}

\usepackage{amsmath}    
\usepackage{graphicx}   
\usepackage{soul,xcolor}
\setstcolor{red}

\usepackage{hyperref}
\usepackage{algorithm}
\usepackage{algpseudocode}
\usepackage{tikz,pgf}
\usepackage{epsfig,amssymb,amsmath,ifthen,comment,mathtools}
\usepackage{color}
\usepackage{array}
\usepackage{pdflscape}
\usepackage[figuresright]{rotating}
\usepackage{xr}

\usetikzlibrary{arrows,shapes.arrows,calc,shapes.geometric,shapes.multipart,  
decorations.pathmorphing,positioning,shapes.swigs}

\def\ci{{\perp\!\!\!\perp}}

\newcommand{\G}{{\mathcal G}}

\newenvironment{prf}{\noindent\textit{Proof:}\begin{mdseries}}{\end{mdseries}{\hfill\scriptsize$\Box$}}

\newtheorem{thm}{Theorem}
\newtheorem{lem}{Lemma}

\newtheorem{rmk}{Remark}

\newtheorem{cor}{Corollary}
\newtheorem{prop}{Proposition}

\DeclareMathOperator{\med}{med}

\DeclareMathOperator{\pa}{pa}

\DeclareMathOperator{\ch}{ch}

\DeclareMathOperator{\pas}{pas}

\DeclareMathOperator{\Ch}{Ch}

\jmlrheading{1}{2021}{1-?}{8/15}{??/??}{robins21}{James M. Robins, Thomas S. Richardson, and Ilya Shpitser}

\begin{document}

\title{An Interventionist Approach to Mediation Analysis}

\author{\name James M.\ Robins \email robins@hsph.harvard.edu\\
	\addr Harvard T. H. Chan School of Public Health\\
	Boston, MA 02115, USA
	\AND
       \name Thomas S.\ Richardson \email thomasr@uw.edu\\
       \addr Department of Statistics\\
       University of Washington\\
	Seatle, WA 98195-4550, USA
       \AND
	\name Ilya Shpitser \email ilyas@cs.jhu.edu\\
       \addr Department of Computer Science\\
	Johns Hopkins University\\
       Baltimore, MD 21218, USA
}

\editor{Anonymous}

\maketitle

\begin{abstract}
Judea Pearl's insight that, when errors are assumed independent, the Pure (aka Natural) Direct Effect (PDE) is non-parametrically identified via the Mediation Formula was ``path-breaking'' in more than one sense.  In the same paper Pearl described a thought-experiment as a way to motivate
the PDE. Analysis of this experiment led Robins and Richardson to a novel way of conceptualizing direct effects in terms of interventions on an expanded graph 
in which treatment is decomposed into multiple separable components. We further develop this novel theory here, showing that it
 provides a self-contained framework for discussing mediation without reference to cross-world (nested) counterfactuals or interventions on the mediator.
The theory preserves the dictum ``no causation without manipulation'' and makes questions of mediation empirically testable in future randomized controlled trials.
Even so, we prove the interventionist and nested counterfactual approaches remain tightly coupled under a Non-Parametric Structural Equation Model except in the presence of a ``recanting witness.''
In fact, our analysis also leads to a simple sound and complete algorithm for determining identification in the (non-interventionist) theory of path-specific counterfactuals.
\end{abstract}

\keywords{causal inference; mediation analysis}

\begin{figure}
	\begin{center}
		\begin{tikzpicture}[>=stealth, node distance=1.2cm]
		\tikzstyle{format} = [draw, thick, circle, minimum size=4.0mm,
		inner sep=1pt]
		\tikzstyle{unode} = [draw, thick, circle, minimum size=1.0mm,
		inner sep=0pt,outer sep=0.9pt]
		\tikzstyle{square} = [draw, very thick, rectangle, minimum size=4mm]

	\begin{scope}[xshift=0cm]
		\path[->,  line width=0.9pt]
		node[format, shape=ellipse] (a) {$A$}
		node[format, shape=ellipse, above right of=a, yshift=-0.3cm, fill=lightgray] (s) {$S$}			
		node[format, shape=ellipse, below right of=a, yshift=+0.3cm] (m) {$M$}
		
		node[format, shape=ellipse, left= of s, xshift=0.25cm, fill=lightgray] (u) {$U$}

		node[format, shape=ellipse, right of=s,xshift=-0.2cm, fill=lightgray] (r) {$R$}
		node[format, shape=ellipse, right of=m,xshift=-0.2cm] (y) {$Y$}

		(a) edge[blue] (s)
		(a) edge[blue] (m)
		(s) edge[blue] (r)
		(m) edge[blue] (r)
		(m) edge[blue] (y)
		(r) edge[blue] (y)
		
		(u) edge[blue, bend left] (r)
		(u) edge[blue, out=270,in=180] (m)

		node[below of=m]{(a)}	
		;
	\end{scope}

	\begin{scope}[xshift=6.1cm]
		\path[->,  line width=0.9pt]
		node[format, shape=ellipse] (a) {$A$}
		node[ shape=ellipse, above right of=a, yshift=-0.3cm] (s) {}			
		node[format, shape=ellipse, below right of=a, yshift=+0.3cm] (m) {$M$}
		
		node[ shape=ellipse, left= of s, xshift=0.25cm] (u) {}

		node[ shape=ellipse, right of=s,xshift=-0.2cm] (r) {}
		node[format, shape=ellipse, right of=m,xshift=-0.2cm] (y) {$Y$}

		(a) edge[blue] (m)
		(m) edge[blue] (y)
		

		(a) edge[blue, bend left] (y)
		(m) edge[<->, bend right, red] (y)

		node[below of=m]{(a$^*$)}	
		;
		\end{scope}
		
	\begin{scope}[yshift=-3.5cm,xshift=-0.6cm]

		\begin{scope}
			\tikzset{line width=0.9pt, inner sep=1.8pt, swig vsplit={gap=6pt, inner line width right=0.3pt}}	
				\node[ xshift=0.0cm, yshift=0.0cm, name=a, shape=swig vsplit]{
        					\nodepart{left}{$A$}
        					\nodepart{right}{$a$} };
		\end{scope}

		\path[->,  line width=0.9pt]
		node[format, shape=ellipse, above right of=a, xshift=0.5cm, yshift=-0.3cm, fill=lightgray] (s) {$S(a)$}			
		;

		\begin{scope}
			\tikzset{line width=0.9pt, inner sep=1.8pt, swig vsplit={gap=6pt, inner line width right=0.3pt}}	
				\node[below right of=a, xshift=0.7cm, yshift=+0.3cm, name=m, shape=swig vsplit]{
        					\nodepart{left}{$M(a)$}
        					\nodepart{right}{$m$} };
		\end{scope}

		\path[->,  line width=0.9pt]

		node[format, shape=ellipse, left= of s, xshift=-0.1cm, fill=lightgray] (u) {$U$}

		node[format, shape=ellipse, right of=s, xshift=0.8cm, fill=lightgray] (r) {$R(a,m)$}
		node[format, shape=ellipse, right of=m, xshift=1.2cm] (y) {$Y(a,m)$}

		(a) edge[blue] (s)
		(a) edge[blue] (m)
		(s) edge[blue] (r)
		(m) edge[blue] (y)
		(r) edge[blue] (y)
		(m) edge[blue, out=20,in=240] (r)

		(u) edge[blue, out=270,in=180] (m)
		
		(u) edge[blue,bend left] (r)

		node[below of=m]{(b)}	
		;
		\end{scope}

	\begin{scope}[yshift=-3.5cm, xshift=5.5cm]

		\begin{scope}
			\tikzset{line width=0.9pt, inner sep=1.8pt, swig vsplit={gap=6pt, inner line width right=0.3pt}}	
				\node[ xshift=0.0cm, yshift=0.0cm, name=a, shape=swig vsplit]{
        					\nodepart{left}{$A$}
        					\nodepart{right}{$a$} };
		\end{scope}

		\path[->,  line width=0.9pt]
		node[ shape=ellipse, above right of=a, xshift=0.5cm, yshift=-0.3cm] (s) {}			
		;

		\begin{scope}
			\tikzset{line width=0.9pt, inner sep=1.8pt, swig vsplit={gap=6pt, inner line width right=0.3pt}}	
				\node[below right of=a, xshift=0.7cm, yshift=+0.3cm, name=m, shape=swig vsplit]{
        					\nodepart{left}{$M(a)$}
        					\nodepart{right}{$m$} };
		\end{scope}

		\path[->,  line width=0.9pt]
		node[ shape=ellipse, right of=s, xshift=0.4cm] (r) {}
		node[format, shape=ellipse, right of=m, xshift=1.2cm] (y) {$Y(a,m)$}

		(a) edge[blue] (m)
		(m) edge[blue] (y)
		
		(a) edge[blue, bend left] (y)
		
		(m) edge[<->, red, out=280, in=260, looseness=0.8] (y)

		node[below of=m]{(b$^*$)}	
		;
		\end{scope}
		
	\begin{scope}[yshift=-7cm,xshift=-0.6cm]

		\begin{scope}
			\tikzset{line width=0.9pt, inner sep=1.8pt, swig vsplit={gap=6pt, inner line width right=0.3pt}}	
				\node[ xshift=0.0cm, yshift=0.0cm, name=a, shape=swig vsplit]{
        					\nodepart{left}{$A$}
        					\nodepart{right}{$a\!=\!0\,$} };
		\end{scope}

		\begin{scope}
			\tikzset{line width=0.9pt, inner sep=1.8pt, swig vsplit={gap=6pt, inner line width right=0.3pt}}	
				\node[below right of=a, xshift=0.7cm, yshift=+0.3cm, name=m, shape=swig vsplit]{
        					\nodepart{left}{$M(0)$}
        					\nodepart{right}{$m$} };
		\end{scope}
		
		\path[->,  line width=0.9pt]
		node[format, shape=ellipse, above right of=a, xshift=0.5cm, yshift=-0.3cm, fill=lightgray] (s) {$S(0)$}			

		node[format, shape=ellipse, left= of s, xshift=-0.15cm, fill=lightgray] (u) {$U$}

		node[format, shape=ellipse, right of=s, xshift=0.8cm, fill=lightgray] (r) {$R(0,m)$}
		node[format, shape=ellipse, right of=m, xshift=1.2cm] (y) {$Y(0,m)$}

		(a) edge[blue] (s)
		(a) edge[blue] (m)
		(s) edge[blue] (r)
		(m) edge[blue] (y)
		(r) edge[blue] (y)
		
		(u) edge[blue, out=220,in=200] (m)
		(m) edge[blue, out=20,in=240] (r)

		node[below of=m]{(c)}	
		;
		\end{scope}
		
	\begin{scope}[yshift=-7cm, xshift=5.5cm]

		\begin{scope}
			\tikzset{line width=0.9pt, inner sep=1.8pt, swig vsplit={gap=6pt, inner line width right=0.3pt}}	
				\node[ xshift=0.0cm, yshift=0.0cm, name=a, shape=swig vsplit]{
        					\nodepart{left}{$A$}
        					\nodepart{right}{$a\!=\!0\,$} };
		\end{scope}

		\begin{scope}
			\tikzset{line width=0.9pt, inner sep=1.8pt, swig vsplit={gap=6pt, inner line width right=0.3pt}}	
				\node[below right of=a, xshift=0.7cm, yshift=+0.3cm, name=m, shape=swig vsplit]{
        					\nodepart{left}{$M(0)$}
        					\nodepart{right}{$m$} };
		\end{scope}
		
		\path[->,  line width=0.9pt]
		node[ shape=ellipse, above right of=a, xshift=0.5cm, yshift=-0.3cm] (s) {}			


		node[ shape=ellipse, right of=s, xshift=0.3cm] (r) {}
		node[format, shape=ellipse, right of=m, xshift=1.2cm] (y) {$Y(0,m)$}

		(a) edge[blue] (m)
		(m) edge[blue] (y)
		
		(a) edge[blue, bend left] (y)

		;
		\end{scope}
		
		\begin{scope}
		
		\begin{scope}
			\tikzset{line width=0.9pt, inner sep=1.8pt, swig vsplit={gap=6pt, inner line width right=0.3pt}}	
				\node[below right of=a, xshift=0.7cm, yshift=+0.3cm, name=m, shape=swig vsplit]{
        					\nodepart{left}{$M(0)$}
        					\nodepart{right}{$m$} };
		\end{scope}
		
		\path[->,  line width=0.9pt]


		node[format, shape=ellipse, right of=m, xshift=1.2cm] (y) {$Y(0,m)$}

		(a) edge[blue] (m)
		(m) edge[blue] (y)
		

		node[below of=m]{(c$^*$)}	
		;
		\end{scope}

	\begin{scope}[yshift=-10.5cm,xshift=-0.6cm]

		\begin{scope}
			\tikzset{line width=0.9pt, inner sep=1.8pt, swig vsplit={gap=6pt, inner line width right=0.3pt}}	
				\node[ xshift=0.0cm, yshift=0.0cm, name=a, shape=swig vsplit]{
        					\nodepart{left}{$A$}
        					\nodepart{right}{$a\!=\!1\,$} };
		\end{scope}

		\path[->,  line width=0.9pt]
		node[format, shape=ellipse, above right of=a, xshift=0.5cm, yshift=-0.3cm, fill=lightgray] (s) {$S(1)$}			
		;
		
		\begin{scope}
			\tikzset{line width=0.9pt, inner sep=1.8pt, swig vsplit={gap=6pt, inner line width right=0.3pt}}	
				\node[below right of=a, xshift=0.7cm, yshift=+0.3cm, name=m, shape=swig vsplit]{
        					\nodepart{left}{$M(1)$}
        					\nodepart{right}{$m$} };
		\end{scope}
		
		\path[->,  line width=0.9pt]
		
		node[format, shape=ellipse, left= of s, xshift=-0.1cm, fill=lightgray] (u) {$U$}

		node[format, shape=ellipse, right of=s, xshift=0.8cm, fill=lightgray] (r) {$R(1,m)$}
		node[format, shape=ellipse, right of=m, xshift=1.2cm] (y) {$Y(1,m)$}

		(a) edge[blue] (s)
		(a) edge[blue] (m)
		(s) edge[blue] (r)
		(m) edge[blue] (y)
		(r) edge[blue] (y)
		
		(u) edge[blue,bend left] (r)
		(m) edge[blue, out=20,in=240] (r)

		node[below of=m]{(d)}	
		;
	\end{scope}

		\begin{scope}[yshift=-10.5cm, xshift=5.5cm]

		\begin{scope}
			\tikzset{line width=0.9pt, inner sep=1.8pt, swig vsplit={gap=6pt, inner line width right=0.3pt}}	
				\node[ xshift=0.0cm, yshift=0.0cm, name=a, shape=swig vsplit]{
        					\nodepart{left}{$A$}
        					\nodepart{right}{$a\!=\!1\,$} };
		\end{scope}
		\begin{scope}
			\tikzset{line width=0.9pt, inner sep=1.8pt, swig vsplit={gap=6pt, inner line width right=0.3pt}}	
				\node[below right of=a, xshift=0.7cm, yshift=+0.3cm, name=m, shape=swig vsplit]{
        					\nodepart{left}{$M(1)$}
        					\nodepart{right}{$m$} };
		\end{scope}
		
		\path[->,  line width=0.9pt]


		node[format, shape=ellipse, right of=m, xshift=1.2cm] (y) {$Y(1,m)$}

		(a) edge[blue] (m)
		(m) edge[blue] (y)
		
		(a) edge[blue, bend left] (y)

		node[below of=m]{(d$^*$)}	
		;
		\end{scope}

		\end{tikzpicture}
		\end{center}
\caption{
(a) A DAG $\cal G$ representing the underlying structure in the combined observational and randomized trial of treatments for river blindness.
Shaded variables are not observed.
Here $A$ indicates a randomized trial ($a=1$) or an observational study ($a=0$);
$M$ is whether the patient received ivermectin; $Y$ is the patient's vision; $S$ indicates access to  a clinic; $R$ is treatment with immuno suppressants; $U$ indicates the inclination of the patient to avail themselves of medical care that is offered. (b) The SWIG ${\cal G}(a,m)$ resulting from $\cal G$; (c) and (d) show SWIGs ${\cal G}(a=0,m)$ and
${\cal G}(a=1,m)$ that incorporate additional context specific causal information. (a$^*$), (b$^*$), (c$^*$), (d$^*$) show the corresponding latent projections.
\label{fig:torpedo3}}
\end{figure}
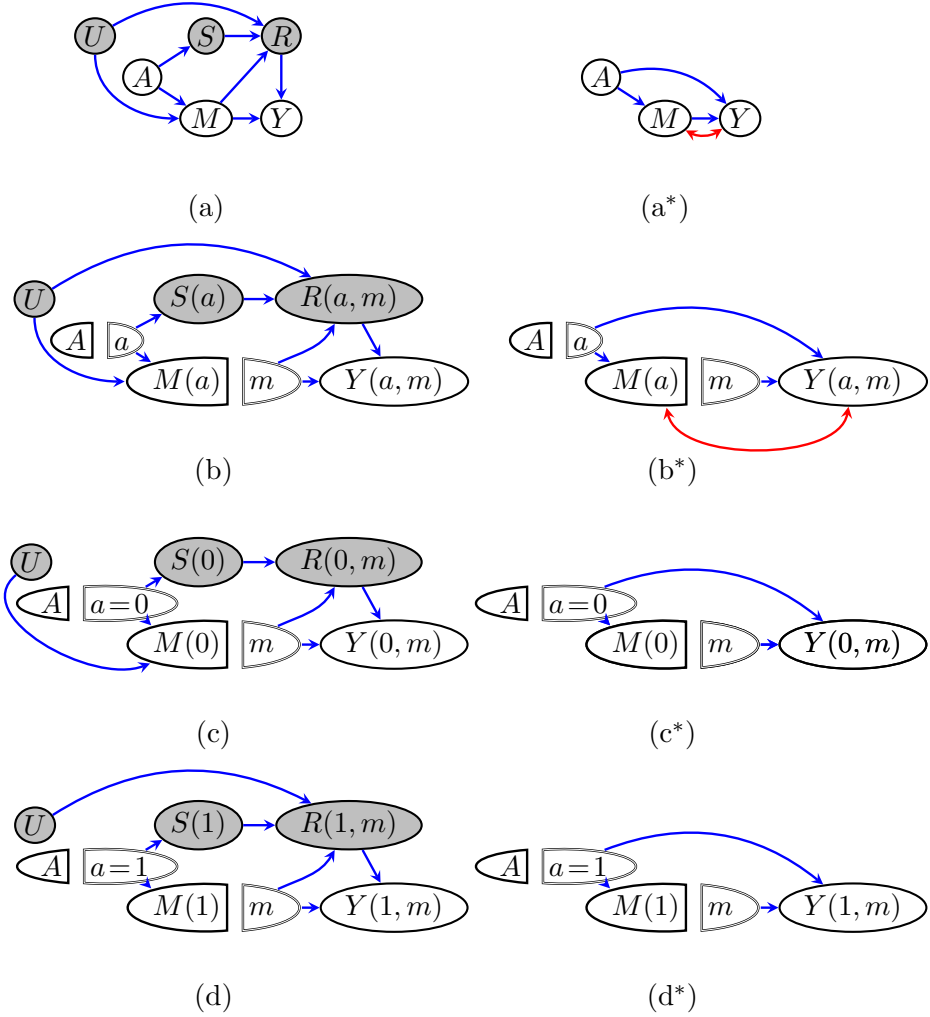

\section{Introduction}
\label{sec:mediation}

In the companion paper \citep{rrs21volume_id_arxiv} we described graphical counterfactual models corresponding
to the finest fully randomized causally interpretable structured tree graph (FFRCISTG) models of \cite{robins86new}. 
Such models correspond to conditional independencies encoded by the d-separation criterion in Single World Intervention Graphs (SWIGs).  We gave a general identification theory for treatment effects in such models with hidden variables that is a generalization of  the theory \citet{tian02on} and \citet{shpitser06id}. We also
described the differences between the FFRCISTG/SWIG model and the Non-Parametric Structural Equation Model with Independent Errors (NPSEM-IE).

Given that a treatment effect of $A$ on $Y$ is established, it is often desirable to try to understand 
the contribution to the total effect of $A$ on $Y$ of the
other variables $M$
that lie on causal pathways from $A$ to $Y$.  
Variables lying on the causal pathways are called mediators of the effect of $A$ on $Y$.
This leads to consideration of direct and indirect effects.
Several different types of direct effect have been considered previously. Most have asked whether ``the outcome ($Y$) would have
been different had cause ($A$) been different, but the level of the mediator ($M$) remained unchanged.'' 
Differences arise regarding what it means to say that $M$ remains unchanged.

In Section \ref{sec:mediation-based-on-m}, we will first review these notions, the assumptions under which they are identified and the extent to which identification claims can be verified (in principle) via an experiment. These considerations will lead to a novel way of conceptualizing 
direct effects introduced in \citep{robins10alternative} and generalized in Section \ref{sec:intervention-mediation-theory}  herein.
This novel interventionist approach does not require counterfactuals defined in terms of the mediator.
We first describe our interventionist approach to causal mediation analysis in the context of direct and indirect effects.
This approach (i) need not assume that the  mediator $M$ has well defined causal effects and/or counterfactuals, but (ii) instead hypothesizes that treatment variable $A$ can be decomposed into multiple separable components each contributing to the overall  effect of treatment,  (iii) preserves the dictum ``no causation without manipulation,'' (iv)  makes questions of mediation empirically testable in future randomized controlled trials, (v) may facilitate communication with subject matter experts,  and (vi) when identified from data, the identifying formulae under an interventionist approach and those obtained from the other (mediator-based) approach are identical; however the causal effects being identified differ since they refer to intervening on different variables. This theory has been extended and applied by others
in the context of mediation in survival analysis \citep{didelez19defining,aalen20timedependent} and recently to competing risks 
\citep{stensrud:separable-effects:2020,stensrud:separable:2020} and interference problems \citep{shpitser17modeling}; see also \citep{lok:2016:organic}.

Finally in section \ref{sec:path-specific}, we extend this approach to arbitrary (identified) path-specific effects.
This allows the (generalized) ID algorithm described in \citep{rrs21volume_id_arxiv} plus one extra step to be used as a sound and complete algorithm for determining the identification of  path-specific distributions; we also provide  a version for handling conditional queries. Finally, we describe the differences between the nested counterfactual approach and the interventionist approach under non-identification.


\section{Approaches To Mediation Based On Counterfactuals Defined In Terms Of The Mediator: The CDE and PDE}\label{sec:mediation-based-on-m}
One approach to having $M$ remain unchanged would be for $M$ to be fixed via an intervention.
On this view both $A$ and $M$ are treatments, and we consider the difference between an intervention setting $A$ to $a$ and $M$ to $m$, versus an intervention setting $A$ to $a'$ and $M$ to $m$.  This leads to the definition of a {\it controlled direct effect} (CDE): 
\begin{equation}\label{eq:cde}
CDE_{a,a'}(m) \equiv \mathbb{E}[Y(a',m) - Y(a,m)],
\end{equation}
where $Y(a,m)$ is the counterfactual response of $Y$ had $A$ and $M$ been set, possibly contrary to fact, to values $a$ and $m$, respectively.

Note that there is a controlled direct effect for every value $m$ of $M$.
Given a causal graph $\cal G$, a straightforward application of graphical causal identification theory\footnote{%
See the companion paper \citep{rrs21volume_id_arxiv} and references therein.}
 determines whether the distribution $P(Y(a,m))$, and thus whether the  $CDE_{a,a'}(m)$ contrast, is identified. 


There are situations in which other notions of direct effect are more natural.  In particular, there are contexts in which we wish to know whether $A$ taking the value of $a'$ (rather than the baseline level $a$) would lead to a change in the value of $Y$ if the effect of $A$  on $M$ were ``blocked.'' Specifically, if the effect on $M$ of $A$ taking {value} $a'$ was blocked so that the mediator $M$ takes the the value $M(a)$ that $M$ would take were $A$ set to the baseline value $a$.

Along these lines, in many contexts we may ask what ``fraction'' of the (total) effect of $A$ on $Y$ may be attributed to a particular causal pathway. For example, consider a randomized controlled trial that investigates the effect of an anti-smoking intervention ($%
A$) on the outcome myocardial infarction (MI) at 2 years ($Y$) among non-hypertensive smokers. For
simplicity, assume everyone in the treatment arm and no one in the
placebo arm quit cigarettes, that all subjects were tested for new-onset
hypertension ($M$) at the end of the first year, and no subject suffered
an MI in the first year. Hence $A$, $M$, and $Y$ occur in that
order. Suppose the trial showed smoking cessation had a beneficial effect on
both hypertension and MI. It is natural to ask: ``What
fraction of the total effect of smoking cessation ($A$) on MI ($Y$) is
via a pathway that does not involve hypertension ($M$)?''

These ideas lead to the {\it pure direct effect}\footnote{Also called the natural direct effect in \citep{pearl01direct}.} defined in \citep{robins92effects},
which in counterfactual notation can be written as follows:
 \begin{align}\label{eq:pde}
 PDE_{a,a'} \equiv \mathbb{E}[Y(a',M(a)) - Y(a,M(a))] = \mathbb{E}[Y(a',M(a)) - Y(a)].
 \end{align}

This is the difference between
two quantities: first, the outcome $Y$ that would result if we set $A$
to $a'$, while ``holding fixed'' $M$ at the value $M(a)$ that
it would have taken had $A$ been $a$; second, the outcome $Y$ that would
result from simply setting $A$ to $a$ (and thus having $M$ again
take the value $M(a)$).  In \citep{robins92effects}, the first was alternately described as the result of setting $A$ to $a'$ on $Y$ when the the effect of $A$ on $M$ is blocked, thereby leaving $M$  unchanged from its reference value $M(a)$.
Thus the Pure Direct Effect interprets
had ``$M$ remained unchanged'' to mean ``had (somehow) $M$ taken the
value that it would have taken had we fixed $A$ to $a$.'' 

As another application of this idea, \citet[p.~131]{pearl09causality} cites the following legal opinion arising in a discrimination case:
{\it
``The central question in any employment discrimination case is whether the employer would have taken the same action had the employee been of a different race (age, sex, religion, national origin etc.) and everything else had been the same.''}
{\small (Carson vs.~Bethlehem Steel Corp., 70 FEP Cases 921, 7th Cir. (1996)).} 

Here $A$ corresponds to membership in a protected class, while $M$ corresponds to criteria, such as qualifications, that are permitted to be considered in such decisions. If the PDE is non-zero then discrimination has taken place.

A notion of the indirect effect may be defined similarly, using the nested counterfactual $Y(a',M(a))$.  On the additive scale, direct and indirect effects may be used to give a decomposition of the average causal effect  \citep{robins92effects}:
\begin{align}
\underbrace{\mathbb{E}[Y(a')] - \mathbb{E}[Y(a)]}_{\text{average causal effect}} &= 
\underbrace{(\mathbb{E}[Y(a')] - \mathbb{E}[Y(a',M(a))])}_{\text{total indirect effect}} +
\underbrace{(\mathbb{E}[Y(a',M(a))] - \mathbb{E}[Y(a)])}_{\text{pure direct effect}}\nonumber\\
&=
\underbrace{(\mathbb{E}[Y(a')] - \mathbb{E}[Y(a,M(a'))])}_{\text{total direct effect}} +
\underbrace{(\mathbb{E}[Y(a,M(a'))] - \mathbb{E}[Y(a)])}_{\text{pure indirect effect}}.\label{eq:tde-intro}
\end{align}

Notice that the PDE depends on $Y(a',M(a))$ --- a variable in which two different levels of $a$ are nested within the counterfactual for $Y$.
Consequently, in contrast to $CDE_{a,a'}$ this counterfactual does not correspond to any experimental intervention on $A$ and $M$.  
This is because in order to know the value $M(a)$ for a unit, it is necessary to set $A$ to $a$, but this then precludes setting $A$ to $a'$. This is a manifestation of the fundamental problem of causal inference: it is not possible for a single unit to receive two different levels of the same treatment at the same time.
For this reason the counterfactual $Y(a',M(a))$ is referred to as a {\it cross-world} counterfactual.\footnote{
Formally, we will say that a counterfactual expression is {\it cross-world} if it involves assigning more than one value to a single index, such as $a$ and $a'$ in $Y(a',M(a))$.}

The differences in the assumptions made by the FFRCISTG and NPSEM-IE models
 lead to quite different identification results for the PDE in the context of the simple DAG shown in Figure~\ref{fig:amyno}~(a). These differences reflect important epistemological distinctions between the two frameworks,  that are described in Sections \ref{subsec:pde-npsem-ie} and \ref{subsec:pde-ffrcistg} below.

\subsection{Two Hypothetical River Blindness Treatment Studies}
\label{ex:riverblindness}

We will use as a running example the following pair of hypothetical studies that are represented together in the single causal graph shown in Figure~\ref{fig:torpedo3}~(a), as described below:
A random sample of individuals in an impoverished medically underserved
catchment area are selected to participate in a double-blind
placebo controlled randomized trial ($A=1$ selected, $A=0$ otherwise) of 
single dose therapy with the drug Ivermectin ($M=1$ received the drug, $M=0$ otherwise)
for the treatment of onchocerciasis (river blindness). The outcome is
diminished vision 9 months
later ($Y=1$ if worse than $20/100$ and $Y=0$ otherwise). All subjects in the trial complied with their assigned therapy. The
trial was motivated in part by the fact that Ivermectin was already being
sold by local shop owners as a cure for river blindness without evidence for
effectiveness or safety in the local population. After the trial finished, an
NGO carried out a retrospective observational cohort study on a random
subset of non-selected subjects ($A=0$), collecting data on $M$ and $Y$. 

In a subset of patients Ivermectin can actually decrease visual acuity.
This occurs when the larvae in the eye killed by Ivermectin  induce an over
vigorous immune response. Because of this side effect, a
clinic available to all trial participants was established to screen for the above side-effect
 and treat with immune suppressive drugs if required ($R=1$ if treated
with immune suppressants, $R=0$ otherwise) to prevent further eye damage. The
 patients not selected for the study had neither access to the clinic nor to immunosuppresive
therapy. Thus letting $S=1$ ($S=0$) denote access (no access) to a clinic, we
have $S=1$ iff $A=1$ and $R=0$ if $A=0$. Finally we let $U$ denote an unmeasured 
variable with $U=1$ and $U=0$ denoting individuals with greater versus lesser
propensity to take medical treatments if offered.   In the unselected
subjects ($A=0$), $U$ is thus positively correlated with $M$. In contrast, owing to
random assignment of $M$, $U$ and $M$ are independent in selected subjects ($A=1$).
On the other hand, in selected subjects, conditional on $M$, $U$ is positively
correlated with $R$ and thus with $Y$; in unselected subjects $U$ and $Y$ are
independent given $M$ because treatment with $R$ is not available. (Here we assume $U$ has
an effect on $Y$ only through the treatments actually received.) Finally, 
records recording which of the selected subjects received immuno-suppressive
therapy in the trial were destroyed in a fire so the data available for analysis were solely 
$(A,M,Y)$, the same variables available in the observational study.

\subsection{The PDE And CDE In The River Blindness Studies}

We next explain the meaning of the CDE and PDE in the context of the Ivermectin studies.
There we suppose the true data-generating process is as described by the {Directed Acyclic Graph (}DAG{)} in {Figure~\ref{fig:torpedo3}~(a)}.
Recall that the corresponding latent projection\footnote{See Section~\protect\ref{subsec:latent} in the companion paper \citep{rrs21volume_id_arxiv} for this definition and other standard graphical definitions.}
is given in Figure~\ref{fig:torpedo3}~(a$^*$)
because in Figure~\ref{fig:torpedo3}~(a) there is a causal path from $A$ to $Y$, and $U$ is a common cause of $M$ and $Y$.
In this context $M(a=0)$ is the ivermectin treatment that the patient would select in the observational study where $A$ is $0$. Likewise $Y(a=1,M(a=0))$ is the patient's outcome
if they received the ivermectin treatment as in the observational study  (where $A$ is $0$) but, as in the randomized study (where $A$ is $1$),
 a clinic ($S=1$) was made available to them.\footnote{%
If patients know that a clinic is available then this may influence their decision to take ivermectin; thus in order for the patient's decision to be as in the observational study, they would need to make decisions relating to ivermectin treatment  before being told that a clinic is available.}
 To see why, note that $Y(a=1,M(a=0))$ corresponds to the effect of setting $a=1$ on $Y$ through all causal pathways ($S\rightarrow R\rightarrow Y$) not passing through ivermectin ($M$) when  $M$ remains at its self selected value  $M(a=0)$ in the observational study.
 Finally $Y(a=0) \equiv Y(a=0,M(a=0))$ is the patient's outcome if they were assigned to the observational study
 and did not have access to the clinic. The PDE is the mean of the difference $Y(a=1,M(a=0)) - Y(a=0,M(a=0))$. 

The $CDE(m=0)$ is the mean effect on $Y$ of having ($s=1$) versus not having ($s=0$)  a clinic available  had no one received ivermectin ($m=0$).\footnote{This is because $Y(a,m)=Y(S(a),m)=Y(s,m)$ since $S(a)=a$.} Likewise, $CDE(m=1)$ is the clinic effect when all subjects received ivermectin ($m=1$). In contrast, as noted above, the PDE is the mean effect of having versus not having a clinic available had subjects chosen ivermectin treatment ($M$) as in the observational study. 

%

\subsubsection*{Identification Of The CDE In The River Blindness Studies}

We next consider whether the  $CDE(m=0)$ and  $CDE(m=1)$ are identified from the 
available data on $A$, $M$ and $Y$.
From examining the latent projection shown in Figure~\ref{fig:torpedo3}~(a$^*$), one would expect
that the CDE is not identified owing to the bidirected edge $M\leftrightarrow Y$.
In particular, identification does not follow from existing methods
such as the {\it do}-calculus, the ID algorithm\footnote{Note that the ID algorithm is complete \citep{shpitser06id} with respect to models defined in terms of the independences holding in a (standard) causal DAG; these independences are not context specific.}
 or the back-door criterion \citep{pearl09causality}, though see \citep{tikka:2019context-specific}.

However, we will now show, perhaps surprisingly, that we do have identification of the CDE.
This is a consequence of context specific independencies.
Although such independencies cannot be represented using standard causal DAGs (or their latent projections) 
an extension to context specific SWIGs due to  \citet{dahabreh2019generalizing} and \citet{sarvet:graphical:2020} makes this possible.

Consider the SWIG ${\cal G}(a,m)$ resulting from a joint intervention setting $A$ to $a$ and $M$ to $m$
as shown in Figure~\ref{fig:torpedo3}~(b) and its latent projection shown in Figure~\ref{fig:torpedo3}~(b$^*$). 
This SWIG ${\cal G}(a,m)$ shown in Figure~\ref{fig:torpedo3}~(b), represents the conditional independence relations that hold for all four of the possible counterfactual distributions ($a,m\in\{0,1\}$) resulting from jointly intervening on $A$ and $M$.


Consider the two interventions setting $a=1$ and $m \in \{ 0, 1 \}$.  These interventions correspond to performing the randomized trial in which treatment $M$ is randomly assigned
and thus $U \,\ci\, M(a=1)$. Consequently, the distribution of the counterfactuals may be represented by the SWIG ${\cal G}(a=1,m)$, shown
in Figure~\ref{fig:torpedo3}~(d), in which the edge $U\rightarrow M(a=1)$ is absent.

Now consider the remaining \st{two} 
intervention\st{s}
 setting $a=0$, that is, assigned to the observational study. 
Recall that the clinic ($S$) and (hence) treatment with immuno suppressants ($R$) are only available to those people who were selected for the trial ($a=1$). 
Thus for all subjects in the observational study ($a=0$), $S(0)=R(0,m)=0$. Consequently, $U \,\ci\, R(a\!=\!0,m) \mid A, {M(a\!=\!0)}$, since the inclination ($U$) of the patient does not affect whether they have access to immunosuppressants.
Hence the distribution of the counterfactuals may be represented by the SWIG ${\cal G}(a=0,m)$, shown
in Figure  \ref{fig:torpedo3}(c), in which the edge $U\rightarrow R(a=0,m)$ is absent.\footnote{Since $S(0)$ and $R(0,m)$ are constants we could also leave out the edges $S(0)\rightarrow R(0,m)$ and $m\rightarrow R(0,m)$ on the same basis, but this would not change our conclusions.}
%
%
%
%


Applying d-separation to the latent projections in Figure~\ref{fig:torpedo3}~(c$^*$) and (d$^*$) we see that\footnote{Recall that when testing d-separation in SWIGs, fixed nodes
such as $a=0$ in Figure~\ref{fig:torpedo3}~(c$^*$) and $a=1$ in Figure~\ref{fig:torpedo3}~(d$^*$) always block paths they occur on (when they are not end points).}
\begin{equation}\label{eq:yam-ind-ma-a2}
Y(a,m) \,\ci\, M(a), A\quad \hbox{ for }a=0,1.
\end{equation}
Consequently, since $Y=Y(a,M(a)=m)$ on the event $A=a$, $M(a)=m$,
\begin{equation}\label{eq:am-intervention-identified2}
p(Y\,|\,A=a,M=m) = p(Y(a,m)), 
\end{equation}
 is identified for both $a=0$ and $a=1$.
Hence the $CDE(m)$ of $A$ and $M$ on $Y$ is identified.

As noted above, the identification (Equation (\ref{eq:am-intervention-identified2})) does not follow from  the DAG in Figure~\ref{fig:torpedo3}~(a), with the latent projection in Figure~\ref{fig:torpedo3}~(a$^*$). However, the  
 independences $U \ci R(a\!=\!0,m) \mid A, M(a\!=\!0)$ and $U {\ci}\, M(a\!=\!1) \mid A$ {\it can} be encoded by and read from the context specific SWIGs shown in Figure~\ref{fig:torpedo3}~(c) and (d), respectively.\footnote{These independences cannot be read from the 
SWIG ${\cal G}(a,m)$ shown in Figure~\ref{fig:torpedo3}~(b) that was constructed from ${\cal G}$. Note that the graph ${\cal G}(a,m)$ contains the union of edges in the two context-specific SWIGs ${\cal G}(a=0,m)$ and ${\cal G}(a=1,m)$ shown in Figure~\ref{fig:torpedo3}~(c),(d) respectively.
${\cal G}(a,m)$ thus represents the (non context specific) conditional independence relations that are common to all four instantiations $a,m\in \{0,1\}$.}

Together with consistency, these independences imply, respectively, $U \ci R \mid A=0$ and $U \ci M \mid A=1$.
Since, in addition to Equation (\ref{eq:yam-ind-ma-a2}), we also have $M(a) \ci A$ for $a=0,1$, it follows that the distribution of the counterfactuals $\{A, M(a), Y(a,m) \hbox{ for all } a,m \}$ obeys the FFRCISTG model associated with the graph shown in Figure~\ref{fig:amyno}~(a) in which there are no bi-directed edges.
Interestingly, we show below in Section \ref{subsec:ffrcistg-not-npsem-ie} that
$M(a\!=\!0)\not\!\!\ci Y(a\!=\!1,m)$.
Consequently,
the distribution of the counterfactuals does not obey the NPSEM-IE associated with Figure~\ref{fig:amyno}~(a),
and thus the PDE is not identified without additional assumptions beyond those of the NPSEM-IE model.\footnote{The distribution does obey the NPSEM-IE (hence also the FFRCISTG) associated with Figure~\ref{fig:torpedo3}~(a$^*$)
which includes the $M\leftrightarrow Y$ edge; see also Footnote \protect\ref{foot:detectable}.}

\subsection{Identification Of The PDE Via The Mediation Formula Under The NPSEM-IE 
For Figure~\protect\ref{fig:amyno}~(a)}\label{subsec:pde-npsem-ie}

We now consider the situation in which, unlike the ivermectin example above, the NPSEM-IE associated with the graph in Figure~\ref{fig:amyno}~(a) holds. In this case the $PDE_{a,a'}$ is identified via the following 
{\emph{mediation formula}}
 \citep{pearl01direct,pearl12causal}:
\begin{align}
{\med}_{a,a'} &\equiv \sum_{m} \left(\mathbb{E}[Y\,|\, m,a] - \mathbb{E}[Y\,|\, m,a']\right) p(m\,|\,a')\\
                    &= \left(\sum_{m} \mathbb{E}[Y\,|\, m,a]p(m\,|\,a')\right) - \mathbb{E}[Y\,|\, a']\label{eq:mediation}
\end{align}

The proof of this result, under the NPSEM-IE is as follows:
\begin{align*}
\MoveEqLeft{p(Y(a,M(a')=m)=y)}\\
 &= \sum_{m} p(Y(a,M(a')=m)=y\,|\,M(a')=m) p(M(a')=m)\\
&= \sum_{m}p(Y(a,m)=y) p(M(a')=m)\\
&= \sum_{m}p(Y=y\,|\,A=a,M=m) p(M=m\,|\,A=a').
\end{align*}
Here the first line follows from elementary probability, the second from the cross-world NPSEM-IE independence: 
\[
Y(a,m) \ci M(a')
\]
which follows from 
(\ref{eqn:mwm}); the third 
follows from the FFRCISTG independence (\ref{eqn:swm}) and thus also from (\ref{eqn:mwm}).
Under the NPSEM-IE associated with Figure~\ref{fig:torpedo}~(a$^*$) identification fails because this  cross world independence does not hold.

\subsection{Partial Identification Of The PDE Under The FFRCISTG For Figure~\protect\ref{fig:amyno}(a)}\label{subsec:pde-ffrcistg}

In contrast, under the less restrictive FFRCISTG model associated with the graph in Figure \ref{fig:amyno} (a) the PDE is not, in general, identified.
This follows from the fact that the second equality in the previous proof relies on the cross-world independence  
$Y(a,m) \ci M(a')$; but the FFRCISTG model does not assume any cross-world independencies.
However, under the FFRCISTG the observed data implies bounds on the PDE. For example, in the case in which $M$ and $Y$ are binary we have the following sharp bounds \citep{robins10alternative}:
\begin{eqnarray*}
\lefteqn{ \max\{ 0, p(M\!=\!0 \;|\; A\!=\!a')+ p(Y\!=\!1 \;|\; A\!=\!a,M\!=\!0) -1\} \;+} \\
\lefteqn{\kern10pt \max\{0, p(M\!=\!1\;|\; A\!=\!a')+ p(Y\!=\!1\;|\; A\!=\!a,M\!=\!1) -1\}
-p(Y\!=\!1 \mid A=a' )} \\[5pt]
&&\kern20pt \leq\;\; PDE_{a,a'}\;\;\leq \\[5pt]
&& {\ \min\{ p(M\!=\!0 \;|\; A\!=\!a'), p(Y\!=\!1\;|\; A\!=\!a,M\!=\!0)\}\;+ } \\
&&{\kern20pt \min\{ p(M\!=\!1 \;|\; A\!=\!a'), p(Y\!=\!1\;|\; A\!=\!a,M\!=\!1)\}  -p(Y\!=\!1 \mid A=a' ).}
\end{eqnarray*}

\noindent{A proof is given in the Appendix.}

\subsection{An Example In Which An FFRCISTG Model Holds, But An NPSEM-IE Does Not}\label{subsec:ffrcistg-not-npsem-ie}

The differing results on identifiability for the PDE in the previous two sections
raise the question as to whether it is most appropriate to adopt the NPSEM-IE (\ref{eqn:mwm}) or FFRCISTG (\ref{eqn:swm}) assumptions in practice.
As shown above, this choice matters since, if one assumes the NPSEM-IE associated with the simple graph in Figure~\ref{fig:amyno}~(a) then one will believe the PDE is point identified, while if one assumes the FFRCISTG associated with Figure~\ref{fig:amyno}~(a) only bounds may be obtained. 


It has been argued that it is hard to conceive of a realistic data-generating process under which the FFRCISTG model holds, but the NPSEM-IE model does not. However, we now show that the river blindness studies 
above provide\st{s} 
a counterexample.

Recall that due to records being destroyed in a fire, only the three variables ($A$,$M$,$Y$) are observed. The first ($A$) is randomly assigned and the analyst assumes (possibly incorrectly) that there is no other variable (either measured or unmeasured) that is a cause of both the mediator $M$ and the final response $Y$.
 This hypothesis would imply the causal structure depicted in Figure~\ref{fig:amyno}~(a). The graph in Figure~\ref{fig:amyno}~(a) would be the latent projection of Figure~\ref{fig:torpedo3}~(a)
if it were assumed incorrectly that there is no unmeasured confounder $U$. Both the FFRCISTG and NPSEM-IE models associated with Figure~\ref{fig:amyno}~(a) imply:

\begin{equation}\label{eq:sw-torpedo}
Y(a,m) \ci M(a), \quad \hbox{ for } a=0,1.
\end{equation}
The NPSEM-IE also implies the cross-world independence:
 \begin{equation}\label{eq:mw-torpedo}
Y(a=1,m) \ci M(a=0).
\end{equation}

We have already shown that the underlying data-generating process in the ivermectin example implies (\ref{eq:sw-torpedo}); see (\ref{eq:yam-ind-ma-a2}) above. We now show that  (\ref{eq:mw-torpedo}) fails to hold for almost all laws corresponding to the NPSEM-IE associated with Figure \ref{fig:torpedo3}(a). This remains true even if we impose, in addition, the context-specific counterfactual independences needed to identify the CDE that are encoded in Figures~\ref{fig:torpedo3}~(c) and (d).

%

\begin{figure}
	\begin{center}
		\begin{tikzpicture}[>=stealth, node distance=1.2cm]
		\tikzstyle{format} = [draw, thick, circle, minimum size=4.0mm,
		inner sep=1pt]
		\tikzstyle{unode} = [draw, thick, circle, minimum size=1.0mm,
		inner sep=0pt,outer sep=0.9pt]
		\tikzstyle{square} = [draw, very thick, rectangle, minimum size=4mm]
	\begin{scope}[yshift=-7cm]

		\begin{scope}
			\tikzset{line width=0.9pt, inner sep=2pt, swig vsplit={gap=6pt, inner line width right=0.3pt}}	
				\node[ xshift=0.0cm, yshift=0.0cm, name=a, shape=swig vsplit]{
        					\nodepart{left}{$A$}
        					\nodepart{right}{$a\!=\!0\,$} };
		\end{scope}
		
		\begin{scope}
			\tikzset{line width=0.9pt, inner sep=2pt, swig vsplit={gap=6pt, inner line width right=0.3pt}}	
				\node[right of=a, xshift=1.8cm, yshift=0.0cm, fill=lightgray, name=s, shape=swig vsplit]{
        					\nodepart{left}{$S(a_0)$}
        					\nodepart{right}{$s\!=\!1\,$} };
		\end{scope}

		\begin{scope}
			\tikzset{line width=0.9pt, inner sep=2pt, swig vsplit={gap=6pt, inner line width right=0.3pt}}	
				\node[below right of=a, xshift=0.7cm, yshift=-0.3cm, name=m, shape=swig vsplit]{
        					\nodepart{left}{$M(a_0)$}
        					\nodepart{right}{$m$} };
		\end{scope}
		
		\path[->,  line width=0.9pt]

		node[format, shape=ellipse, left= of a, xshift=0.8cm, fill=lightgray] (u) {$U$}

		node[format, shape=ellipse, right of=s, xshift=1.8cm, fill=lightgray] (r) {$R(s_1,m)$}
		node[format, shape=ellipse, right of=m, xshift=3.2cm] (y) {$Y(s_1,m)$}

		(a) edge[blue] (s)
		(a) edge[blue] (m)
		(s) edge[blue] (r)
		(m) edge[blue] (y)
		(r) edge[blue] (y)
		
		(u) edge[blue,out=30,in=160] (r)
		(u) edge[blue, out=300,in=170] (m)
		(m) edge[blue, out=20,in=200] (r)

		;
		\end{scope}
		\end{tikzpicture}
		\end{center}	
\caption{The SWIG $\G(a=0,s=1,m)$ associated with the DAG shown in Figure \protect\ref{fig:torpedo3}(a);
here $a_0$ and $s_1$ are short for $a=0$ and $s=1$.
\label{fig:g-of-asm}}
\end{figure}
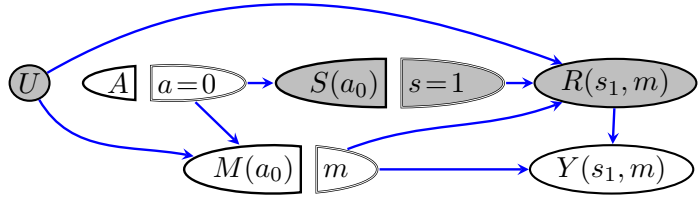

To see this, first consider the SWIG $\G(a=0,s=1,m)$ shown in Figure \ref{fig:g-of-asm} that is constructed from the DAG in Figure \ref{fig:torpedo3}(a).
There is a d-connecting path $Y(s=1,m)$ to $M(a=0)$, namely  $M(a=0) \leftarrow U \rightarrow R(s=1,m) \rightarrow Y(s=1,m)$.
Consequently, for almost all distributions in the FFRCISTG model, it holds that $M(a=0) \not\!\!\!\ci\; Y(s=1,m)$.\footnote{This is also true if we 
assume the NPSEM-IE associated with the graph in Figure~\ref{fig:torpedo3}~(a).} Further, we have:
\begin{equation*}
Y(s=1,m) = Y(S(a=1),m) = Y(a=1, m),
\end{equation*}
where in the first equality we use that, due to determinism, $S(a=1)=1$ for all individuals; the second follows from recursive substitution.\footnote{See Equation (\ref{eqn:rec-sub}) in \protect\citet{rrs21volume_id_arxiv}.}
Hence (\ref{eq:mw-torpedo}) does not hold\footnote{Intuitively, this should not be surprising since $U$ is a `common cause' of $M(a=0)$ and $Y(a=1,m)$
in that $U\rightarrow M(a=0)$ in Fig.~\protect\ref{fig:torpedo3}(c) while there is a path $U\rightarrow R(a=1,m) \rightarrow Y(a=1,m)$ in 
Fig.~\protect\ref{fig:torpedo3}(d).} from which it follows that the PDE is not identified; see Appendix, Section~\ref{subsec:pde-not-identified-river-blindness}.
Consequently the distribution of the counterfactuals $\{A, M(a), Y(a,m)\; \hbox{ for all }a,m\}$ is not in the NPSEM-IE model associated with Figure~\ref{fig:amyno}~(a).
Interestingly we see that by considering a SWIG with interventions on three variables $A$, $S$ and $M$ we have shown that the FFRCISTG model plus recursive substitution and some determinism can be used to prove that a cross world independence fails to hold; see Footnote \ref{foot:detectable} and Appendix, Section~\ref{subsec:detectable-confounding}.

Finally we note that in the ivermectin example the Acyclic Directed Mixed Graph (ADMG)\footnote{See Section \ref{subsec:graphs} in \citep{rrs21volume_id_arxiv}.} with fewest edges (over $A$,  $M$, $Y$) that represents the distribution of the counterfactuals $\{A, M(a), Y(a,m)$ for all  $a,m\}$
is  Figure~\ref{fig:amyno}~(a)  under the FFRCISTG. In contrast, the minimal ADMG that represents this distribution under the NPSEM-IE is the graph with an additional bi-directed confounding arc shown in Figure~\ref{fig:torpedo3}~(a$^*$). This has the following interesting consequence: typically people may make a statement such as ``Figure~\ref{fig:amyno}~(a) is {\em the} true causal graph.'' However, we now see that this statement does not have a truth-value without clarifying whether we are referring to the NPSEM-IE or the FFRCISTG as
 the true underlying counterfactual model.

On the other hand, one might prefer to replace  ``Figure~\ref{fig:amyno}~(a)'' with ``Figure~\ref{fig:torpedo3}~(a$^*$)''  in the statement above, since then the modified statement holds for both counterfactual models;\footnote{Since the distribution of $\{A, M(a), Y(a,m)$
 for all  $a,m\}$ obeys the FFRCISTG corresponding to Figure~\ref{fig:amyno}~(a), which is a subgraph of Figure~\ref{fig:torpedo3}~(a$^*$), 
 the distribution also obeys the FFRCISTG corresponding to this latter graph.}
furthermore,  Figure~\ref{fig:torpedo3}~(a$^*$) accurately indicates that there is confounding for the PDE. However, the disadvantage of this choice is that the inclusion of the bi-directed edge $M\!\leftrightarrow\!Y$ does not reveal that the CDE is identified via $E[Y(a,m)\,|\, A=a, M=m] = E[Y\,|\, A=a,M=m]$. 



\subsection{Testable Versus Untestable Assumptions And Identifiability}\label{subsec:testable-untestable}

Given a graph such as Figure~\ref{fig:amyno}~(a), in principle, there is an empirical test of the FFRCISTG model.\footnote{Formally, this requires that one can observe the ``natural'' value of a variable prior to intervention.} However, there is no additional empirical test (on the variables in the graph) for the extra assumptions made by the corresponding NPSEM-IE, which are required to identify the PDE.
Consequently, there is a qualitative distinction in the testability of the identification assumptions for these two contrasts.

In more detail, identification of the $CDE(m)$, $m=0,1$ is, in principle, subject to direct empirical test:
one conducts a four armed randomized experiment on subjects drawn from the same population, in which both $A$ and $M$ are randomly assigned to their four possible joint values. If for any $(a,m)$ the distribution in the four-arm $(A,M)$ randomized trial 
$p(Y(a,m)=y)$\footnote{Here and throughout this section, we will use $p(A,M,Y)$ to denote the observed distribution in which only $A$ is randomized;
we will use $\{p(Y(a,m))$ for $a,m\in \{0,1\}$ to indicate the four distributions that {\em would} be observed if {\em both} $A$ and $M$ were to be randomized
in a four arm trial. This is because in such a trial in the arm in which people are assigned to $A=a$, $M=m$ we would observe the counterfactual $Y(a,m)$.}
 differs from the conditional distribution in the two-armed trial  $p(Y(a)=y\,|\,M(a)=m)=p(Y=y\,|\,M=m,A=a)$ in which only $A$ was randomized, then we may infer that an unmeasured common cause was present between $M$ and $Y$ and hence it was incorrect to postulate the causal DAG in Figure~\ref{fig:amyno}~(a),
regardless of whether we are considering the FFRCISTG or NPSEM-IE models associated with this graph.

In contrast, whether the PDE equals the mediation formula (\ref{eq:mediation}) cannot be empirically tested using data on $(A,M,Y)$. 
Even if we can directly manipulate $M$ (in addition to $A$), there is no experiment involving $A$ and $M$ such that  the resulting contrast corresponds to the PDE. This is for the following reason: to observe, for a given subject, the cross-world counterfactual $Y(a=1,M(a=0))$ that occurs in the PDE, one would need to {\emph{first}} assign them to $a=0$ and record $M(a=0)$, and {\emph{then}} perform a {\emph{second}} experiment (on the same subject) in which they are assigned to $a=1$ and the recorded value $M(a=0)$ from the {first} experiment.  However, this is usually not possible for the simple reason that having assigned the patient to $a=0$ in the {first} experiment, precludes {\emph{subsequently}}
assigning them to $a=1$, except in the rare circumstances where a valid cross-over trial is feasible.


These considerations are particularly relevant in a setting such as the ivermectin example,  where as shown above, the distribution over the counterfactual variables $\{A, M(a), Y(a,m)$ for all  $a,m\}$ obeys the FFRCISTG but not the NPSEM-IE model associated with the causal DAG in Figure~\ref{fig:amyno}~(a).
In particular, an analyst who was unaware of the variables $U$, $R$, $S$ in Figure~\ref{fig:torpedo3}~(a) and posited the model in Figure~\ref{fig:amyno}~(a) would find no evidence of confounding between $M$ and $Y$ even if they were to subsequently perform a four-arm $(A,M)$ randomized trial.

In summary, the PDE identification via the mediation formula (\ref{eq:mediation}) requires not only that there be no {\it detectable} single-world confounding between $M$ and $Y$ (as assumed by the FFRCISTG), but, in addition, that undetectable cross-world confounding also be absent.%
\footnote{If, after carrying out an experiment in which $A$ and $M$ are randomly assigned, it is observed that $p(Y(a,m))$ and $p(Y\,|\,A=a,M=m)$ are statistically indistinguishable, then this would almost certainly increase  the probability that a Bayesian would assign to cross-world independence holding. 
However, the ivermectin example shows the importance of the investigator -- Bayesian or not -- thinking carefully about the underlying data-generating mechanism.
}
Consequently, as with the ivermectin example, it is possible for the mediation formula to give an inconsistent estimate of the PDE, yet for this to be undetectable given any randomized experiment that could be performed using the variables $A$, $M$, $Y$ on the graph in Figure~\ref{fig:amyno}~(a).

\section{Interventionist Theory Of Mediation}\label{sec:intervention-mediation-theory}


The above considerations motivate a theory of mediation based on interventions on sub-components of treatment,
rather than on the mediator.

\subsection{Interventional Interpretation Of The PDE Under An Expanded Graph}

As described above, the counterfactual $E[Y(a=1, M(a=0))]$ and thus the PDE cannot be empirically tested by any 
intervention on the variables on the graph.
However, curiously, Pearl has often argued that the PDE 
is a causal contrast of substantive and public-health importance by
offering examples along the following lines. 

\subsubsection*{Example: Nicotine-Free Cigarette}

Consider the example discussed in Section~\ref{sec:mediation-based-on-m} where we have data from a randomized
smoking-cessation trial. We have data available on smoking
status $A$, hypertensive status $M$ $6$ months after randomization and myocardial infarction (MI)
status $Y$ at one year. 
 
Following a similar argument given in \citep{pearl01direct}\footnote{\cite{pearl01direct}, Section 2 considers a similar example, but where $A$ is a drug, $M$ is aspirin taken to mitigate side-effects, and $Y$ is the final outcome.} to motivate the PDE, suppose that nicotine-free cigarettes will be newly available starting a year from now.
The substantive goal is to use the already collected data from the smoking cessation trial to estimate
the difference two years from now in the incidence of MI if all smokers were to change to nicotine-free cigarettes when they become available (in a year) compared to the incidence if all smokers were to stop smoking altogether (in a year).

Further suppose it is believed that the entire effect of nicotine
on MI is through its effect on hypertensive status, while the non-nicotine
toxins in cigarettes have no effect on hypertension and that there do not 
exist unmeasured confounders for the effect of hypertension on MI.

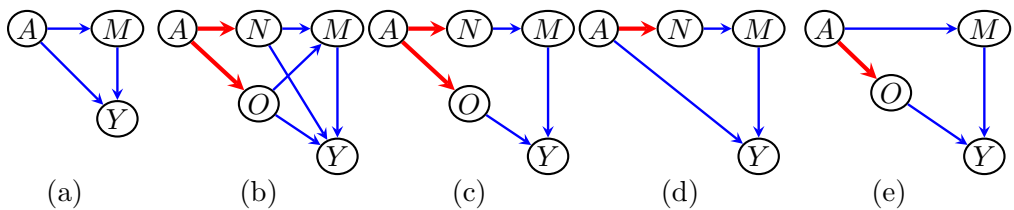
\begin{figure}
\begin{center}
		\begin{tikzpicture}[>=stealth, node distance=1.2cm]
		\tikzstyle{format} = [draw, thick, circle, minimum size=4.0mm,
		inner sep=1pt]
		\tikzstyle{unode} = [draw, thick, circle, minimum size=1.0mm,
		inner sep=0pt,outer sep=0.9pt]
		\tikzstyle{square} = [draw, very thick, rectangle, minimum size=4mm]
\begin{scope}[xshift=-1cm]
		\begin{scope}
		\path[->,  line width=0.9pt]
		node[format, shape=ellipse] (a) {$A$}
		node[format, shape=ellipse, right of=a] (m) {$M$}			
		node[format, shape=ellipse, below of=m] (y) {$Y$}
		(a) edge[blue] (m)
		(m) edge[blue] (y)
		(a) edge[blue] (y)
		node[below of=a, yshift=-1cm, xshift=0.5cm] (l) {(a)}
		;
		\end{scope}

		\begin{scope}[xshift=2cm]
		\path[->,  line width=0.9pt]
		node[format, shape=ellipse] (a) {$A$}
		node[format, shape=ellipse, right =0.5cm of a] (n) {$N$}
		node[format, shape=ellipse, below = 0.5cm of n] (o) {$O$}
		node[format, shape=ellipse, right =1.5cm of a] (m) {$M$}
		node[format, shape=ellipse, below=1.2cm of m] (y) {$Y$}			
		(a) edge[red,ultra thick] (n)
		(a) edge[red, ultra thick] (o)
		(n) edge[blue] (m)
		(o) edge[blue] (y)
		(m) edge[blue] (y)
		(n) edge[blue] (y)
		(o) edge[blue] (m)
		node[below of=n, yshift=-1cm, xshift=0.0cm] (l) {(b)}
		;
		\end{scope}
		
		\begin{scope}[xshift=4.8cm]
		\path[->,  line width=0.9pt]
		node[format, shape=ellipse] (a) {$A$}
		node[format, shape=ellipse, right =0.5cm of a] (n) {$N$}
		node[format, shape=ellipse, below = 0.5cm of n] (o) {$O$}
		node[format, shape=ellipse, right =1.5cm of a] (m) {$M$}
		node[format, shape=ellipse, below=1.2cm of m] (y) {$Y$}			
		(a) edge[red,ultra thick] (n)
		(a) edge[red, ultra thick] (o)
		(n) edge[blue] (m)
		(o) edge[blue] (y)
		(m) edge[blue] (y)
		node[below of=n, yshift=-1cm, xshift=0.0cm] (l) {(c)}
		;
		\end{scope}
		\begin{scope}[xshift=7.6cm]
		\path[->,  line width=0.9pt]
		node[format, shape=ellipse] (a) {$A$}
		node[format, shape=ellipse, right =0.5cm of a] (n) {$N$}
		node[format, shape=ellipse, right =1.5cm of a] (m) {$M$}
		node[format, shape=ellipse, below=1.2cm of m] (y) {$Y$}			
		(a) edge[red,ultra thick] (n)
		(n) edge[blue] (m)
		(a) edge[blue] (y)
		(m) edge[blue] (y)
		node[below of=n, yshift=-1cm, xshift=0.0cm] (c) {(d)}
		;
		\end{scope}
		\begin{scope}[xshift=10.6cm]
		\path[->,  line width=0.9pt]
		node[format, shape=ellipse] (a) {$A$}
		node[format, shape=ellipse, below right = 0.5cm and 0.5cm of a] (o) {$O$}
		node[format, shape=ellipse, right =1.5cm of a] (m) {$M$}
		node[format, shape=ellipse, below=1.2cm of m] (y) {$Y$}			
		(a) edge[red, ultra thick] (o)
		(a) edge[blue] (m)
		(o) edge[blue] (y)
		(m) edge[blue] (y)
		node[right = 2cm of c] (l) {(e)}
		;
		\end{scope}
				
%
%
%
%
\end{scope}
		\end{tikzpicture}
\end{center}
\caption{
(a) A simple DAG ${\cal G}$ 
representing a causal model with a treatment $A$, 
a mediator $M$ and a response $Y$.
(b) An expanded causal model where $N$ and $O$ are a decomposition of $A$;
 $N$ and $O$ are, respectively, the nicotine and non-nicotine components of tobacco. Thicker red edges indicate deterministic relations.
(c) An expanded version ${\cal G}^{\text{ex}}$ of the DAG ${\cal G}$ in (a), and an edge subgraph of the DAG in (b), where $N$ does not cause $Y$ directly,
and $O$ does not cause $M$ directly. 
In this graph the direct and indirect effects of $A$ on $Y$ may be defined via interventions on  $N$ and $O$.
(d) Special case of (c) in which $A$ plays the role of $O$;
(e) Special case of (c) in which $A$ plays the role of $N$.
}
\label{fig:amyno}
\end{figure}

In this context a researcher following the approach that has been advocated by Pearl may postulate that the smoking cessation trial is represented by the 
NPSEM-IE model associated with Figure~\ref{fig:amyno}~(a). 
Under these assumptions, the MI incidence in smokers of
cigarettes free of nicotine would be $E\left[ Y( a=1,M(a=0)) \right]$
since the hypertensive status
of smokers of nicotine-free cigarettes will equal their hypertensive status
under non-exposure to cigarettes. Thus 
$E\left[ Y( a=1,M(a=0)) \right] $ is precisely the 
incidence of MI in smokers 2 years  from now were all smokers to change
to nicotine-free cigarettes a year from now, and thus the PDE:
\begin{equation}
PDE = E\left[ Y( a=1,M(a=0)) \right] - E\left[ Y( a=0,M(a=0)) \right]
\end{equation}
 is the causal contrast of interest.

Given the assumption of an NPSEM-IE  it follows that $E\left[ Y( a=1,M(a=0)) \right]$ equals $\sum_{m}E\left[
Y\,|\,A=1,M=m\right] p(m\,|\,A=0)$, and therefore the PDE is identified from the mediation formula applied to the data from the smoking cessation trial.

What is interesting about Pearl's motivation is that to argue for the
substantive importance of the parameter $E[ Y( a=1,M( a=0))]$, 
he tells a story about the effect of a manipulation --- a manipulation
that makes no reference to $M$ at all. Rather, in this context, the manipulation is to
intervene to eliminate the nicotine component of cigarettes.

The most direct representation of this story is provided by the
{\it expanded} DAG ${\cal G}^{\text{ex}}$ in Figure~\ref{fig:amyno}~(c)
where $N$ is a binary variable representing {\bf n}icotine exposure, $O$ is a
binary variable representing exposure to the {\bf o}ther non-nicotine components of a
cigarette. 
The bolded arrows from $A$ to $N$ and $O$ indicate deterministic relationships: $N(a) = O(a) = a$. 
This is because in the factual data (with probability one) either one smokes
normal cigarettes so $A=N=O=1$ or one is a nonsmoker (i.e. ex-smoker) and $A=N=O=0$. 

This expanded graph now provides a simple interventional interpretation of the PDE. 
The researcher's assumption of the NPSEM-IE associated with Figure~\ref{fig:amyno}~(a) together with the existence
of the variables $N$ and $O$ and associated counterfactuals 
imply the nested counterfactual $Y(a=1,M(a=0))$ is equal to the simple counterfactual $Y(n=0,o=1)$, the outcome had we intervened to expose all subjects to the non-nicotine components, but not to the nicotine components.\footnote{Notice that here we show that the ``cross-world'' counterfactual $Y(a=1,M(a=0))$ defined in
the DAG in Figure~\ref{fig:amyno}~(a) is equal (as a random variable) to the ``non-cross cross-world'' counterfactual  $Y(n=0,o=1)$ associated with the DAG in Figure~\ref{fig:amyno}~(c); see Section~\ref{sec:back-to-pde} for further discussion.}
It follows that
\begin{equation}\label{eq:no-pde}
PDE = E[Y(n=0,o=1)] - E[Y(n=0,o=0)].
\end{equation}
Furthermore these assumptions imply that the graph in Figure~\ref{fig:amyno}~(c) is an FFRCISTG. It then follows from Proposition \ref{prop:swig-pva-identified} in \citep{rrs21volume_id_arxiv} that  
\begin{equation}\label{eq:two-to-four}
E[Y(n=0,o=1)] = \sum_{m} E[Y\,|\, O=1, M=m] p(m\,|\,N=0),
\end{equation}
where the RHS is the g-formula. Thus $E[Y(n=0,o=1)]$ is identified  provided that the terms  in the g-formula are functions of the distribution of the factuals $p(A,N,O,M,Y)$. 
Since in the factual data now available there is  no subject with $N=0$ and $O=1${,} positivity fails and one might suppose 
that $E[Y(n=0,o=1)]$ is not identified, but in fact it is. To see this,
note that the event $\{O=1,M=m\}$ is equal to the event $\{A=1,M=m\}$ and similarly the event $\{N=0,M=m\}$ is equal to the event $\{A=0,M=m\}$ owing to determinism.
Thus by substituting these events we conclude that
\begin{equation}\label{eq:two-to-fourb}
E[Y(n=0,o=1)] = \sum_{m} E[Y\,|\, A=1, M=m] p(m\,|\,A=0),
\end{equation}
which coincides with (one of the terms in) the mediation formula.\footnote{%
\label{foot:tautology-jamie}Implications of Determinism: The  FFRCISTG models associated with Figures~\ref{fig:amyno}~(b) and (c) both imply the factuals $(A,N,O,M,Y)$  
factor with respect to the corresponding graph. Now in Figure \protect\ref{fig:amyno}(c) note (i) $N$ is d-separated  from $Y$ given $\{M,O\}$  and (ii) $O$ is d-separated from $M$ given $N$ which imply $N\ci Y\mid O,M$ and $O \ci M\mid N$, respectively.
In contrast on Figure~\protect\ref{fig:amyno}~(b) neither of the above d-separations hold. Yet, since by the determinism $A=N=O$ as random variables,
both independencies also hold for the DAG in Figure~\protect\ref{fig:amyno}~(b).\par
\quad Furthermore, $E[Y(n=0,o=1)]$ would be identified by the g-formula:%
\[
 \sum_{m} E[Y\,|\, N=0, O=1, M=m] p(m\,|\,N=0,O=1)
\]
were the formula a function of the factual distribution. However, it is not; the event $\{N=0,O=1\}$ has probability zero due to determinism. Notwithstanding this, a naive application of the above independences might lead one to conclude that this g-formula is equal to the RHS of Equation
(\protect\ref{eq:two-to-four}) and thus is identified by Equation (\protect\ref{eq:two-to-fourb}). The error in this argument is that $O \ci M \mid N$
does not imply $p(M=m\,|\,N=0)$ equals $p(M=m\,|\,N=0,O=1)$ when the event $\{O=1,N=0\}$ has probability zero, since the latter is not well-defined.\par
\quad Note that, in the presence of determinism, we have two DAGs with different adjacencies that represent the same set of factual distributions.
The counterfactuals corresponding to the DAGs in Figures~\ref{fig:amyno}~(b) and (c) do not represent the same set of counterfactual distributions; see Footnote \protect\ref{fig:footnote-amyno-swig-graph}.
}


For a researcher following \citet{pearl01direct}, having at the outset assumed an NPSEM-IE associated with the DAG in
Figure~\ref{fig:amyno}~(a), the story involving $N$ and $O$ does not contribute to identification; rather,
it  served only to show that the PDE encodes a substantively important parameter.

However, from the FFRCISTG point of view, the story  not only provides an interventional interpretation of the PDE
but in addition makes the PDE identifiable with the mediation formula being the identifying formula.
Furthermore, Pearl's story makes refutable the claim that the PDE is identified by the mediation formula. Specifically, when nicotine-free cigarettes become available, Pearl's claim can be tested by an intervention that forces a random sample of the population to smoke nicotine-free cigarettes; if the mean of $Y$  under this intervention differs from the RHS of (\ref{eq:two-to-four}), Pearl's claim is falsified.
As this refers to an an actual intervention, the variables $(N,O)$ are not simply formal constructions.
Without knowledge of the substantive meaning of  $N$ and $O$ the trial in which $N$ is set to $0$ and $O$ is set to $1$ is not possible,
even in principle.
See \citep{robins10alternative} and \citep{stensrud:separable-effects:2020} for discussion of substantive considerations regarding whether variables $N$ and $O$ exist
that satisfy the no direct effect assumptions of Figure~\ref{fig:amyno}~(c).\footnote{One can trivially construct artificial variables $N^*\equiv A$ and $O^*\equiv A$ such that the (degenerate) joint distribution of the factuals
$p(A,N^*,O^*,M,Y)$ will factorize according to the DAG in Figure~\protect\ref{fig:amyno}~(c)  and thus satisfy $M \ci O^*,A \mid N^*$ and  $Y\ci N^* \mid A, O^*,M$.
However, these latter independencies are tautologies owing to determinism and thus do not establish e.g.~$Y(o,m) \ci M(n)$ 
as required by the FFRCISTG associated with  Figure~\ref{fig:amyno}~(c).\label{foot:trivial-n-and-o}}

\begin{rmk}
It should be noted that, in the following sense, it is sufficient to find one of the variables $N$ or $O$:
Specifically, if we have a well-defined intervention $N$, satisfying:
\begin{itemize}
\item[(n1)] $N(a) = a$ so that $N=A$ in the observed data;
\item[(n2)] $N$ has no direct effect on $Y$ relative to $A$ and $M$, so $Y(a,n,m) = Y(a,m)$;
\item[(n3)] $A$ has no direct effect on $M$ relative to $N$ so $M(a,n)=M(n)$,
\end{itemize}
\noindent then $A$ will satisfy the conditions for $O$; see Figure \ref{fig:amyno} (d).
Conversely, if there is a well-defined intervention $O$ such that:
\begin{itemize}
\item[(o1)] $O(a) = a$ so that $O=A$ in the observed data;
\item[(o2)] $A$ has no direct effect on $Y$ relative to $O$ and $M$, so $Y(a,o,m) = Y(o,m)$;
\item[(o3)] $O$ has no direct effect on $M$ relative to $A$ so $M(a,o)=M(a)$,
\end{itemize}
then $A$ will satisfy the conditions for $N$; see Figure~\ref{fig:amyno}~(e).
 
 We use all three variables $(A,N,O)$ in our subsequent development since this choice is symmetric in $N$ and $O$, covers both cases and more closely aligns with the original motivating Nicotine intervention.
 \label{rmk:part-split}
\end{rmk}

Lastly, note that, as in the ivermectin example, the existence of the interventions $N$, $O$ satisfying the no direct effect conditions do not imply that the mediation formula identifies the PDE, because confounding between $M$ and $Y$ may still be present.
\footnote{\label{foot:detectable}
An attentive reader might wonder how it is that the mediation formula fails to identify $P(Y(n,o))$ in the ivermectin example (with $A$ as ``$N$'' and $S$ as ``$O$''; see Remark 1 above).
As noted earlier, the counterfactual variables $A,M(a),Y(a,m)$ follow the conditional independences
implied by the FFRCISTG model associated with the DAG in Figure~\ref{fig:amyno}~(a) (in which there is no bi-directed arc between $M$ and $Y$);
see the end of Section \ref{subsec:ffrcistg-not-npsem-ie}. Under this FFRCISTG model, the absence of the $M\leftrightarrow Y$ edge implies that there is no confounding between $M$ and $Y$ that
is {\it detectable} via interventions on $A$ and $M$, since $p(Y(a,m))=p(Y\,|\,A\!=\!a,M\!=\!m)$.\par
%
However, in the ivermectin example, the expanded set of counterfactual variables $A, O, M(a),Y(a,o)$ do {\it not} follow the FFRCISTG model 
corresponding to the DAG in Figure~\ref{fig:amyno}~(e). To see this, consider performing, in addition to the randomized trial where $a=1$ and $s=1$, 
an intervention setting $A$ to $0$ and $O\equiv S$ to $1$, which corresponds to an observational study but with clinics and immunosuppressants.
Notice that the confounding variable $U$ in Figure~\ref{fig:torpedo3}~(a) now becomes detectable in the following sense:
If $U$ were not present in Figure~\ref{fig:g-of-asm2}~(a) then $p(Y(a\!=\!0,s=1)\,|\,M(a\!=\!0)) = p(Y(a\!=\!1,s=1)\,|\,M(a\!=\!1))$.
This follows from Rule 3 of the po-calculus \citep{malinsky19po,rrs21volume_id_arxiv} applied 
to the SWIG 
${\cal G}(a,s)$ derived from Figure~\ref{fig:amyno}~(e) after replacing ``$O$'' with $S$.
However, we show in the Appendix, Section~\ref{subsec:detectable-confounding} that in this example, if $U$ is present
then this equality does not hold in general.
%
%
Consequently, the latent projection over the variables  $A, O\equiv S, M,Y$ includes an $M\leftrightarrow Y$ edge; see also the last paragraph of 
Section~\protect\ref{subsec:testable-untestable}.
}


%

\subsection{Direct And Indirect Effects Via The Expanded Graph}\label{subsec:separability}

In this section we formally introduce the interventionist theory of mediation first introduced in \citep{robins10alternative} and greatly generalized herein.
We do so by continuing with the {nicotine example}.
Recall that $N$ and $O$ were substantively meaningful variables and 
the goal was to use data from a smoking cessation trial to estimate $E[Y(n\!=\!0,o\!=\!1)]$, the incidence of MI through year 2 if all smokers were to change to nicotine-free cigarettes at one year. 
As noted above, this policy intervention was used by Pearl to motivate consideration of the PDE. However, given the public health importance of the policy question, one could instead focus directly on estimating the effect of the proposed 
substantive intervention given data on $(A,M,Y)$ without regard to whether it is equal to the PDE.

In fact, there are many situations where mediation analysis is applied, in which interventions on the putative mediator $M$ are not well-defined.
Consequently, substantive researchers may not wish to make reference to the corresponding counterfactuals,\footnote{We regard the existence of interventions as necessary for counterfactuals to be well-defined, but \citet{pearl18does,pearl19on} and others may take a different view.}
 regardless of whether they may be formally constructed.  For such researchers, the PDE parameter may not be substantively meaningful. 
 Fortunately, the interventionist theory described herein, in contrast to Pearl's approach, does not require reference to counterfactuals indexed by $m$,  such as $Y(a,m)$.%
 \footnote{The $CDE(m)$ and PDE are defined in terms of such counterfactuals.}


%

As noted, the interventionist theory only requires that $N$, $O$, and interventions on them are substantively meaningful.
This is a major advantage since it
 makes it  straightforward to discuss with subject matter experts, for example\st{,} 
 physicians, experiments that would shed light
 on causal pathways; this is a property not shared by the PDE.

Up to this point we have motivated our interventionist theory, based on the expanded graph, as providing an empiricist foundation for the existing 
mediation theory that is based on cross-world (nested) counterfactuals. However, the interventionist theory can be viewed as 
 autonomous,\footnote{Following \citet[Section 6.54]{wittgenstein-1922}, we may view the theory based on nested counterfactuals as a ``ladder'' that we climbed
to reach our empiricist theory, and now having done so, we may choose to ``kick it away.''} providing a self-contained framework for discussing mediation without 
reference to cross-world (i.e.~nested) counterfactuals. In this section, we adopt this viewpoint. 
However we will see that we can prove the two theories are tightly coupled in certain settings; see Section~\ref{sec:back-to-pde}.
\footnote{In related work, \cite{lok:2016:organic} has developed an interventional approach to mediation that also does not require interventions on the mediator in order to be well-defined. Lok introduces a notion of an ``organic intervention'' ($I=1$) that is required to satisfy certain conditions. Lok then defines notions of direct and indirect effects in term of such organic interventions. Our interventional definitions introduced here are similar in spirit to Lok's conditions.

However, in order to capture certain aspects of the concept of direct and indirect effects, we, unlike Lok, also require that the variables ($N$ and $O$) defining our additional interventions (e.g. on $N$) be equal to $A$ in the observed data. Among other things, this ensures that $N(a)=O(a)=a$ and hence $M(a) = M(n=a,o=a)$ and $Y(a) = Y(n=a,o=a)$.

In contrast, an organic intervention could change the mechanism by which the mediator is produced so that the relevant counterfactual random variables for the mediator under the organic intervention ($M(a=0,i=1)$) do not correspond to those in the absence of the organic intervention $M(a=1)$, although they have the same distribution.  See \cite{robins:semantics}, Section 3 for additional discussion in terms of blocking paths.}

Concretely, consider the DAG associated with Figure \ref{fig:amyno} (b). Unlike the 
model associated with Figure \ref{fig:amyno} (a), which involves counterfactuals $M(a)$ and $Y(a)$, the expanded FFRCISTG model
associated with Figure \ref{fig:amyno} (b) involves counterfactuals $M(n,o)$ and $Y(n,o)$.
Taking the interventionist view as primitive, in what follows, we will discuss counterfactuals, such as $Y(n=0,o=1)$ that, without further assumptions, are defined solely within this larger expanded model.

\subsection*{Identification Of Four Arms From Two}


%
%

%
%
%
%

For the purposes of our development, consider the following three datasets all derived from the same distribution $p$ over the one step ahead counterfactuals
in the FFRCISTG model associated with the graph in Figure \ref{fig:amyno}(b) 
\begin{itemize}
\item[(i)] The original observed data from the trial in which $A$ was randomized, namely $A$, $M$, $Y$;
\item[(ii)] Data from a putative four arm ($N,O$) randomized trial; the data  in each arm $(n,o) \in \{0,1\}^2$ corresponds to $M(n,o)$, $Y(n,o)$;
\item[(iii)] A dataset obtained from the four arm ($N,O$) trial (ii) by restricting to the two arms in which $n=o$.
\end{itemize}

Note that in dataset (i) among people with $A=a$ we observe $N=O=a$, owing to determinism; hence we observe $M(n\!=\!a,o\!=\!a)$ and $Y(n\!=\!a,o\!=\!a)$ on this event.\footnote{Note that the assumption that $p(M(a)) = p(M(n=a,o=a))$ and $p(Y(a) | M(a) ) = p(Y(n=a,o=a) | M(n=a))$.  This assumption is subject to empirical test by examining whether the distribution from (i) and (iii) are the same.  This corresponds to the six-arm trial described by \cite{stensrud2020conditional}.  The distribution of (i) and (iii) could differ when, for example, the treatment A contains additional sub-components that are present in neither $N$ nor $O$.\label{ftn:six-arm}}
By randomization of $A$, it follows that for $a\in \{0,1\}$,
\begin{align*}
p(M=m,Y=y\,|\,A=a) &=  p(M(n\!=\!a, o\!=a)=m,Y(n\!=\!a,o\!=a)=y\,|\,A=a)\\
&= p(M(n\!=\!a,o\!=a)=m,Y(n\!=\!a,o\!=a)=y)
\end{align*}
 which is the distribution of individuals in dataset (iii). 
We conclude that the distribution of the  data in (iii) is identified from the observed data (i).
Thus our goal becomes the identification of $E[Y(n,o)]$ for $n\neq o$ from data on $M(n,o)$ and $Y(n,o)$ in the two arms with $n=o$.
Motivated by the nomenclature of \cite{stensrud:separable-effects:2020} when this identification is possible we will say that the effects of $N$ and $O$ on $M$ and $Y$ are {\em separable}. The following proposition provides sufficient conditions.






\subsubsection*{Identifying The Results Of A Future Four Arm Study From A Current Two Arm Study}

\begin{prop}\label{prop:2-to-4}
 If for some $x \in \{0,1\}$ the following two conditions hold: 
\begin{align}
p(M( n\!=\!x,o\!=\!0) =m) &=p(M(n\!=\!x,o\!=\!1) =m),\label{eq:conditions-for-4-to-2b}\\[4pt]
p(Y( n\!=\!1,o\!=\!x^*) &=y \,|\, M( n\!=\!1,o\!=\!x^*) =m)\nonumber \\
 &= p(Y( n\!=\!0,o\!=\!x^*)=y \,|\, M( n\!=\!0,o\!=\!x^*) =m),\label{eq:conditions-for-4-to-2a} 
\end{align}
where $x^*=1-x$, then:
\begin{align}
 \MoveEqLeft{p(M(n\!=\!x,o\!=\!x^*)=m, Y(n\!=\!x,o\!=\!x^*)=y)}\label{eq:mediation-again}\\
 &= p(Y( n\!=\!x^*,o\!=\!x^*)=y \,|\, M(n\!=\!x^*,o\!=\!x^*) =m) p( M( n\!=\!x,o\!=\!x) =m).\nonumber
\end{align}
\end{prop}
Note that by consistency and randomization of $A$ in dataset (i) the RHS of Equation (\ref{eq:mediation-again}), when summed over $m$ is simply the first expression in the mediation formula (\ref{eq:mediation}). Hence under Equations (\ref{eq:conditions-for-4-to-2a}) and (\ref{eq:conditions-for-4-to-2b}) $E[Y(n,0)]$ is identified from data set (iii) by the mediation formula.
\medskip

\noindent {\it Proof:}
\begin{align*} 
\MoveEqLeft[4]{p(M(n\!=\!x,o\!=\!x^*), Y(n\!=\!x,o\!=\!x^*))}\\
 &=
p(Y(n\!=\!x,o\!=\!x^*)\,|\,M(n\!=\!x,o\!=\!x^*) )p(M(n\!=\!x,o\!=\!x^*))\\
&=p(Y(n\!=\!x^*,o\!=\!x^*)\,|\,M(n\!=\!x^*,o\!=\!x^*) )p(M(n\!=\!x,o\!=\!x)).
\end{align*}

\bigskip

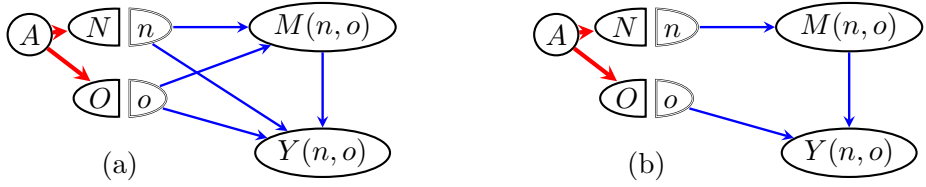
\begin{figure}
\begin{center}
		\begin{tikzpicture}[>=stealth, node distance=1.2cm]
		\tikzstyle{format} = [draw, thick, circle, minimum size=4.0mm,
		inner sep=1pt]
		\tikzstyle{unode} = [draw, thick, circle, minimum size=1.0mm,
		inner sep=0pt,outer sep=0.9pt]
		\tikzstyle{square} = [draw, very thick, rectangle, minimum size=4mm]
		\begin{scope}
		\path[->,  line width=0.9pt]
		node[format,  inner sep=1.8pt, shape=ellipse] (a) {$A$}
		node[format, xshift=0cm, yshift=0.1cm, inner sep=2.4pt, right of = a,
		          name=n, shape=swig vsplit, swig vsplit={gap=4pt, inner line width right=0.3pt}]{
        					\nodepart{left}{$N$}
        					\nodepart{right}{$n$}}
		node[format, xshift=0cm, yshift=0.1cm, inner sep=2.4pt, below=0.5cm of n,
		          name=o, shape=swig vsplit, swig vsplit={gap=4pt, inner line width right=0.3pt}]{
        					\nodepart{left}{$O$}
        					\nodepart{right}{$o$}}
		node[format, shape=ellipse, right =1cm of n] (m) {$M(n,o)$}
		node[format, shape=ellipse, below=1cm of m] (y) {$Y(n,o)$}			
		(a) edge[red,ultra thick] (n)
		(a) edge[red, ultra thick] (o)
		(n) edge[blue] (m)
		(o) edge[blue] (y)
		(n) edge[blue] (y)
		(o) edge[blue] (m)
		(m) edge[blue] (y)
		node[below of=o, yshift=0.3cm, xshift=0.0cm] (l) {(a)}
		;
		\end{scope}
	\begin{scope}[xshift=7cm]
		\path[->,  line width=0.9pt]
		node[format,  inner sep=1.8pt, shape=ellipse] (a) {$A$}
		node[format, xshift=0cm, yshift=0.1cm, inner sep=2.4pt, right of = a,
		          name=n, shape=swig vsplit, swig vsplit={gap=4pt, inner line width right=0.3pt}]{
        					\nodepart{left}{$N$}
        					\nodepart{right}{$n$}}
		node[format, xshift=0cm, yshift=0.1cm, inner sep=2.4pt, below=0.5cm of n,
		          name=o, shape=swig vsplit, swig vsplit={gap=4pt, inner line width right=0.3pt}]{
        					\nodepart{left}{$O$}
        					\nodepart{right}{$o$}}
		node[format, shape=ellipse, right =1cm of n] (m) {$M(n,o)$}
		node[format, shape=ellipse, below=1cm of m] (y) {$Y(n,o)$}			
		(a) edge[red,ultra thick] (n)
		(a) edge[red, ultra thick] (o)
		(n) edge[blue] (m)
		(o) edge[blue] (y)
		(m) edge[blue] (y)
		node[below of=o, yshift=0.3cm, xshift=0.0cm] (l) {(b)}
		;
		\end{scope}
		\end{tikzpicture}
\end{center}
\caption{(a),(b) SWIGs derived from the corresponding expanded DAGs $\G^{ex}$ shown in Figures \protect\ref{fig:amyno}(b),(c) respectively.
In the SWIG shown in (b), the mediator is labeled $M(n,o)$ to indicate that this is a population FFRCISTG for $G^{ex}$, which does not assume
the absence of individual level direct effects.}\label{fig:amyno-swig}
\end{figure}

The constraints in Equations (\ref{eq:conditions-for-4-to-2a}) and (\ref{eq:conditions-for-4-to-2b}) are implied by the SWIG given in Figure \ref{fig:amyno-swig}(b) with
treatments $n$ and $o$ over the random variables variables $N$, $O$, $M(n,o)$ and $Y(n,o)$.\footnote{%
Thus although, as noted previously in footnote \protect\ref{foot:tautology-jamie}, under determinism, the DAGs in Figure{s} \ref{fig:amyno}(b),(c) imply the same conditional independence relations on $p(A,M,Y,N,O)$, they lead to different counterfactual models, since the
constraints {given by Equations} (\protect\ref{eq:conditions-for-4-to-2a}) and (\ref{eq:conditions-for-4-to-2b}) are implied by the SWIG in Figure \protect\ref{fig:amyno-swig}(b),
but not the SWIG in Figure \protect\ref{fig:amyno-swig}(a).\label{fig:footnote-amyno-swig-graph}
}
Note that  $Y(n,o)$ is d-separated from the fixed node $n$ given
$M(n,0)$, which implies under the SWIG global Markov property
that the distribution $p(Y(n,o)\,|\,M(n,o))$ does not
depend on $n$, which is equivalent to (\ref{eq:conditions-for-4-to-2a}).

Similarly, since $M(n,o)$ is d-separated from the fixed node $o$, $p(M(n,o))$ does not depend on $o$. Thus 
we see that the constraints (\ref{eq:conditions-for-4-to-2b}) are implied by the FFRCISTG model.\footnote{As noted in Remark \ref{rmk:part-split}, in some settings $A$ can play the role of $N$ or $O$.}

Thus the identifiability result in Proposition \ref{prop:2-to-4} follows from the fact in the SWIG  in Figure \ref{fig:amyno-swig}(b),
there is no variable whose conditional distributional distribution, given its (random) parents, depends on both $n$ and $o$.


Proposition \ref{prop:2-to-4} above, and indeed all the results in the remainder of this subsection,
holds, when counterfactuals indexed by the mediator $m$ are not well-defined.
Recall that 
\cite{robins86new,robins10alternative} develop an  FFRCISTG model
in which only interventions on a subset of variables are considered well-defined.\\


\noindent{\bf Remarks:} 
\begin{itemize}
\item[1.] The reader may wonder why in this SWIG we have labeled $M$ with
$(n,o)$, rather than simply with $(n)$.  This is 
to emphasize that in this subsection our results do not require the assumption that missing arrows on a SWIG  imply the absence of the associated  direct effect for all individuals. Rather in this sub-section we only impose the weaker assumption that any SWIG is a ``population causal graph.''
A population causal graph \cite[Section 7]{thomas13swig} assumes the distribution of the variables on the graph factor according to the graph, but does not impose the assumption that a missing arrow implies no individual level effects. Thus the variable $M(n,o)$ need  not equal the variable $M(n)=M(n,O)$ and thus $M(n,o)$  cannot be labelled as  $M(n)$.
That is, in the underlying FFRCISTG model associated with the SWIG the one step ahead counterfactuals $M(n,o)$  depend on both $n$ and $o$.

In other words, this population FFRCISTG
contains the counterfactual variables present in the SWIG shown in Figure~\ref{fig:amyno-swig}~(a) that results from splitting $N$ and $O$ in the graph shown in Figure~\ref{fig:amyno}~(b).
It does not correspond to an NPSEM associated with the 
graph in Figure~\ref{fig:amyno}~(c) since that NPSEM assumes well-defined counterfactuals intervening on $M$ and also assumes that $M(n,o) = M(n)$; see also Footnotes \ref{foot:weaker-swigs}, \ref{foot:r3}, and Section \ref{sec:weaker} in \citep{rrs21volume_id_arxiv}.
In particular, under the SWIG derived from the population graph the constraints {in Equations} (\ref{eq:conditions-for-4-to-2a})
and (\ref{eq:conditions-for-4-to-2b}) correspond to the absence of the edges $n\rightarrow Y(n,o)$ and $o\rightarrow M(n,o)$ respectively. 

\item[2.] Since Equations (\ref{eq:conditions-for-4-to-2a}) and (\ref{eq:conditions-for-4-to-2b}) are restrictions on the distribution of the counterfactuals in Figure \ref{fig:amyno-swig}(b), there exist consistent tests of the conditions  (\ref{eq:conditions-for-4-to-2a}) and (\ref{eq:conditions-for-4-to-2b}) given the data from (ii). These conditions cannot be tested given only the data (iii) (or equivalently (i)).\footnote{We note that consistent tests of individual level no direct effect conditions such as $Y(n,o) = Y(o)$ do not exist since
it is possible that $E[Y(n,o)] = E[Y(n,o')]$ and yet $Y(n,o) \neq Y(n,o')$ (as random variables); but see also Footnote \ref{ftn:six-arm}.}

\item[3.] We have seen that under the FFRCISTG associated with Figure \ref{fig:amyno-swig}(b), 
 the distribution of $Y(n,o)$ for all four arms is identified from the two arms in which $n=o$; thus
the structure of this SWIG is sufficient for this identification. However, this structure is also ``necessary'' in that
 it is the only population SWIG over $M(n,o)$ and $Y(n,o)$ 
 where this identification
 is possible.\footnote{Here we are ignoring the structure relating the random variables $N$ and $O$.}
 To see this, first note that if $Y(n,o)$ depends on both $n$ and $o$ then  $n$ and $o$ are both ancestors of $Y$.
By Proposition \ref{prop:2-to-4-general} below, if $n$ and $o$ are both parents of $Y$, then identifiability fails to hold.
Given that we only have one other measured variable then this implies that we must have
one fixed node that is a parent of $M$ (and $M$ in turn a parent of $Y$); the other fixed node is then a parent of $Y$.
If any other edges are present between $\{n,o\}$ and $\{M(n,o),Y(n,o)\}$ then again by Proposition \ref{prop:2-to-4-general}, 
the conditional distributions will not be identifiable.
Likewise, if there is an unmeasured confounder between $M$ and $Y$ then $Y(n,o)$ will not be d-separated from
$n$ given $M(n,o)$.
\end{itemize}

\medskip

\noindent We have the following more general result.

\begin{prop}\label{prop:2-to-4-general}
Assume the distribution of the variables on an unexpanded DAG $\G$ is positive. 
Under an FFRCISTG corresponding to the population SWIG $\G^{ex}(n,o)$ there is no vertex that has both $n$ and $o$ as parents if and only if the joint distributions $p(V(n\!=\!x,o\!=\!x^*))$ for $x \neq x^*$ are identified from the counterfactual distributions $p(V(n\!=\!x, o\!=\!x))$. Further, 
since $P(V(n\!=\!x,o\!=\!x)) = P(V(a\!=\!x))$, also from the distribution of the variables in $\G$.\footnote{Note that the result here holds because the DAG $\G$ here does not contain hidden variables.} 
\end{prop}

{\it Proof:} Consider the g-formula of Proposition \ref{prop:swig-pva-identified} in \citep{rrs21volume_id_arxiv} applied to the graph $\G^{ex}$ under an intervention on $n$ and $o$. This formula is a function of the joint distribution of the observables if and only if none of the terms in the g-formula have both $N$ and $O$ in the conditioning event. Note that each term in the g-formula is the conditional distribution of a variable given its parents on the population SWIG $\G^{ex}(n,o)$.
The requirement here that there is no vertex that is a child of both $n$ and $o$ is directly analogous to the ``no recanting witness'' condition in theory developed by \cite{chen05ijcai}.

\medskip
Consider the following examples from \citep{robins10alternative}.
Suppose our original causal graph $\G$ in Figure~\ref{fig:amyno}~(a) for the cigarette cessation trial was incorrect and the correct causal graph is shown in Figure~\ref{fig:amyl}~(a). There exist three possible $(N,O)$ elaborations of this graph, which are shown in 
Figure~\ref{fig:anomlpath}. These represent different causal theories about the causal effect of the treatment variables $N$, $O$  on $L$, $M$ and $Y$. Figure~\ref{fig:ymlno-swig} shows the corresponding population SWIGs.
Under the SWIGs in Figure{s}~\ref{fig:ymlno-swig}~(a), (b) the distribution of the four arms $Y(n,o)$ is identified given the distributions
$Y(n\!=\!x,o\!=\!x)$:
\begin{align}
p(Y(x,x^{*})) &= \sum_{m,l} p(Y(x^{*},x^{*})\,|\, M(x^{*},x^{*}), L(x^{*},x^{*})) p(M(x,x)\,|\, L(x,x))p(L(x,x))\notag\\
&= \sum_{m,l} p(Y\,|\,m, l, a=x^{*}) p(m\,|\, l,a=x)p(l\,|\,a=x) \label{eq:swig-id-result-a}
\end{align}
and
\begin{align}
p(Y(x,x^{*})) &= \sum_{m,l} p(Y(x^{*},x^{*})\,|\, M(x^{*},x^{*}), L(x^{*},x^{*})) p(M(x,x)\,|\, L(x,x)) p(L(x^{*},x^{*}))\notag\\
&= \sum_{m,l} p(Y\,|\,m, l, a=x^{*}) p(m\,|\, l,a={x})p(l\,|\,a=x^*)\label{eq:swig-id-result-b} 
\end{align}    
respectively, where here we are using $Y(i,j)$ to denote $Y(n\!=\!i,o\!=\!j)$. Note that the identifying formulae are different.
See \citep{stensrud:separable:2020} for generalizations of these results.

In contrast, under the SWIG in Figure \ref{fig:ymlno-swig}(c), $p(Y(n,o))$ for $n\neq o$ is not identified from the data on the two arms with $n=o$ 
(equivalently the observed data) because in $\G^{ex}$ $L$ has both $N$
and $O$ as parents; hence the term $p(l\mid n=x,o=x^*)$ in the g-formula for $p(Y(x,x^*))$ is not a function of the observed data. 

Given that $N$ and $O$ are real interventions at most one of the expanded causal graphs shown in Figure 
\ref{fig:anomlpath} can represent the true causal structure. If, in the future, we obtain data from a four arm $(N,O)$ trial  we can test between the three competing theories associated with these expanded graphs.\footnote{Note however, that if the results from the four arm trial do not correspond
to the identifying formulae obtained from either Figure~\ref{fig:ymlno-swig}~(a) or (b), then although it is possible that the DAG in 
Figure~\ref{fig:ymlno-swig}~(c) holds, it is also possible that the true graph could correspond to Figure~\ref{fig:ymlno-swig}~(a) or (b) with an added unmeasured
confounder between any pair of the variables $L$, $M$ and $Y$.}

%


\medskip

\begin{figure}
	\begin{center}
		\begin{tikzpicture}[>=stealth, node distance=1.2cm]
		\tikzstyle{format} = [draw, thick, circle, minimum size=4.0mm,
		inner sep=1.8pt]
		\tikzstyle{unode} = [draw, thick, circle, minimum size=1.0mm,
		inner sep=0pt,outer sep=0.9pt]
		\tikzstyle{square} = [draw, very thick, rectangle, minimum size=4mm]
		\begin{scope}[xshift=0cm]
		\path[->,  line width=0.9pt]
		node[format, shape=ellipse] (a) {$A$}
		node[format, shape=ellipse, right of=a] (m) {$M$}			
		node[format, shape=ellipse, below of=m] (y) {$Y$}
		node[format, shape=ellipse, right =1cm of m] (l) {$L$}
		node[below=1.3cm of a]{(a)}	
		(a) edge[blue] (m)
		(m) edge[blue] (y)
		(a) edge[blue] (y)
		(a) edge[blue,out=30,in=150] (l)
		(l) edge[blue] (m)
		(l) edge[blue] (y)
		;
		\end{scope}
\begin{scope}[xshift=6cm]
		\path[->,  line width=0.9pt]
		node[format, shape=ellipse] (a) {$A$}
		node[format, shape=ellipse, right of=a] (m) {$M$}			
		node[format, shape=ellipse, below of=m] (y) {$Y$}
		node[format, shape=ellipse, fill=lightgray, right =1cm of m] (l) {$H$}
		node[below=1.3cm of a]{(b)}	
		(a) edge[blue] (m)
		(m) edge[blue] (y)
		(a) edge[blue] (y)
		(l) edge[blue] (m)
		(l) edge[blue] (y)
		;
		\end{scope}
		\end{tikzpicture}
		\end{center}
\caption{
 (a) DAG containing an observed common cause $L$ of the mediator $M$ and outcome $Y$
that is also caused by $A$; (b) DAG containing an unobserved common cause $H$ of the mediator $M$ and outcome $Y$.
\label{fig:amyl}}
\end{figure}
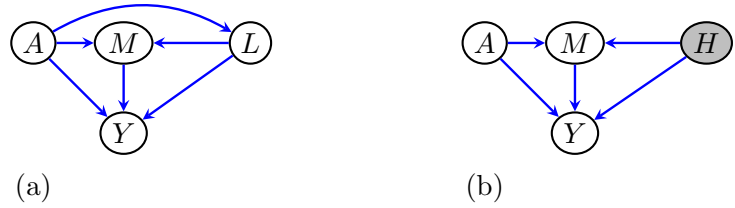




\begin{figure}
\begin{center}
\begin{tikzpicture}[>=stealth, node distance=1.2cm]
\tikzstyle{format} = [draw, thick, circle, minimum size=4.0mm,
		inner sep=1pt]
		\tikzstyle{unode} = [draw, thick, circle, minimum size=1.0mm,
		inner sep=0pt,outer sep=0.9pt]
		\tikzstyle{square} = [draw, very thick, rectangle, minimum size=4mm]
\begin{scope}
		\path[->,  line width=0.9pt]
		node[format, shape=ellipse] (a) {$A$}
		node[format, shape=ellipse, right =0.5cm of a] (n) {$N$}
		node[format, shape=ellipse, below = 0.5cm of n] (o) {$O$}
		node[format, shape=ellipse, right =2cm of a] (m) {$M$}
		node[format, shape=ellipse, right =1cm of m] (l) {$L$}
		node[format, shape=ellipse, below=1cm of m] (y) {$Y$}	
		node[below=0.3cm of y]{(a)}			
		(a) edge[red,ultra thick] (n)
		(a) edge[red, ultra thick] (o)
		(n) edge[blue] (m)
		(n) edge[blue,out=30,in=150] (l)
		(o) edge[blue] (y)
		(m) edge[blue] (y)
		(l) edge[blue] (m)
		(l) edge[blue] (y)
			;
\end{scope}
\begin{scope}[xshift=6cm]
\path[->,  line width=0.9pt]
		node[format, shape=ellipse] (a) {$A$}
		node[format, shape=ellipse, right =0.5cm of a] (n) {$N$}
		node[format, shape=ellipse, below = 0.5cm of n] (o) {$O$}
		node[format, shape=ellipse, right =2cm of a] (m) {$M$}
		node[format, shape=ellipse, right =1cm of m] (l) {$L$}
		node[format, shape=ellipse, below=1cm of m] (y) {$Y$}	
		node[below=0.3cm of y]{(b)}			
		(a) edge[red,ultra thick] (n)
		(a) edge[red, ultra thick] (o)
		(n) edge[blue] (m)
		(o) edge[blue,out=300,in=280,looseness=1.3] (l)
		(o) edge[blue] (y)
		(m) edge[blue] (y)
		(l) edge[blue] (m)
		(l) edge[blue] (y)
			;
\end{scope}
\begin{scope}[yshift=-4cm,xshift=3cm]
\path[->,  line width=0.9pt]
		node[format, shape=ellipse] (a) {$A$}
		node[format, shape=ellipse, right =0.5cm of a] (n) {$N$}
		node[format, shape=ellipse, below = 0.5cm of n] (o) {$O$}
		node[format, shape=ellipse, right =2cm of a] (m) {$M$}
		node[format, shape=ellipse, right =1cm of m] (l) {$L$}
		node[format, shape=ellipse, below=1cm of m] (y) {$Y$}	
		node[below=0.3cm of y]{(c)}			
		(a) edge[red,ultra thick] (n)
		(a) edge[red, ultra thick] (o)
		(n) edge[blue] (m)
		(n) edge[blue,out=30,in=150] (l)
		(o) edge[blue,out=300,in=280,looseness=1.3] (l)
		(o) edge[blue] (y)
		(m) edge[blue] (y)
		(l) edge[blue] (m)
		(l) edge[blue] (y)
			;
\end{scope}
\end{tikzpicture}
\end{center}
\caption{Elaborations of the graph in Figure \protect\ref{fig:amyl} (a),
with additional edges.
These represent different causal theories about the causal effect of the treatment variables $N$, $O$  on $L$, $M$, $Y$.
As before, the thicker red edges indicate deterministic relations.}
\label{fig:anomlpath}
\end{figure}
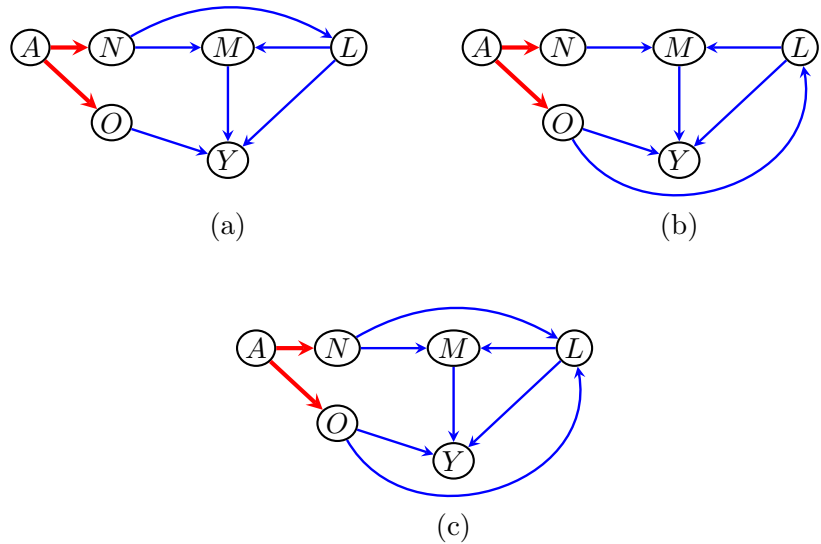

\begin{figure}
\begin{center}
\begin{tikzpicture}[>=stealth, node distance=1.2cm]
\tikzstyle{format} = [draw, thick, circle, minimum size=4.0mm,
		inner sep=1.0pt]
		\tikzstyle{unode} = [draw, thick, circle, minimum size=1.0mm,
		inner sep=0pt,outer sep=0.9pt]
		\tikzstyle{square} = [draw, very thick, rectangle, minimum size=4mm]
\begin{scope}
		\path[->,  line width=0.9pt]
		node[format, shape=ellipse] (a) {$A$}
		node[format, xshift=0cm, yshift=0.1cm, inner sep=2.4pt, above right=0.5mm and 5.0mm of a,
		          name=n, shape=swig vsplit, swig vsplit={gap=4pt, inner line width right=0.3pt}]{
        					\nodepart{left}{$N$}
        					\nodepart{right}{$n$}}
		node[format, xshift=0cm, yshift=0.1cm, inner sep=2.4pt, below=0.5cm of n,
		          name=o, shape=swig vsplit, swig vsplit={gap=4pt, inner line width right=0.3pt}]{
        					\nodepart{left}{$O$}
        					\nodepart{right}{$o$}}
		node[format, shape=ellipse, right =0.5cm of n] (m) {$M(n,o)$}
		node[format, shape=ellipse, right =0.5cm of m] (l) {$L(n,o)$}
		node[format, shape=ellipse, below=0.5cm of m] (y) {$Y(n,o)$}	
		node[below=0.3cm of y]{(a)}			
		(a) edge[red,ultra thick] (n)
		(a) edge[red, ultra thick] (o)
		(n) edge[blue] (m)
		(n) edge[blue,out=30,in=150] (l)
		(o) edge[blue] (y)
		(m) edge[blue] (y)
		(l) edge[blue] (m)
		(l) edge[blue] (y)
		;
		\end{scope}
\begin{scope}[xshift=7cm]
\path[->,  line width=0.9pt]
		node[format, shape=ellipse] (a) {$A$}
		node[format, xshift=0cm, yshift=0.1cm, inner sep=2.4pt, above right=0.5mm and 5.0mm of a,
		          name=n, shape=swig vsplit, swig vsplit={gap=4pt, inner line width right=0.3pt}]{
        					\nodepart{left}{$N$}
        					\nodepart{right}{$n$}}
		node[format, xshift=0cm, yshift=0.1cm, inner sep=2.4pt, below=0.5cm of n,
		          name=o, shape=swig vsplit, swig vsplit={gap=4pt, inner line width right=0.3pt}]{
        					\nodepart{left}{$O$}
        					\nodepart{right}{$o$}}
		node[format, shape=ellipse, right =0.5cm of n] (m) {$M(n,o)$}
		node[format, shape=ellipse, right =0.5cm of m] (l) {$L(n,o)$}
		node[format, shape=ellipse, below=0.5cm of m] (y) {$Y(n,o)$}	
		node[below=0.3cm of y]{(b)}				
		(a) edge[red,ultra thick] (n)
		(a) edge[red, ultra thick] (o)
		(n) edge[blue] (m)
		(o) edge[blue,out=300,in=240,looseness=1] (l)
		(o) edge[blue] (y)
		(m) edge[blue] (y)
		(l) edge[blue] (m)
		(l) edge[blue] (y)
			;
\end{scope}
\begin{scope}[xshift=3cm, yshift=-3cm]
\path[->,  line width=0.9pt]
		node[format, shape=ellipse] (a) {$A$}
		node[format, xshift=0cm, yshift=0.1cm, inner sep=2.4pt, above right=0.5mm and 5.0mm of a,
		          name=n, shape=swig vsplit, swig vsplit={gap=4pt, inner line width right=0.3pt}]{
        					\nodepart{left}{$N$}
        					\nodepart{right}{$n$}}
		node[format, xshift=0cm, yshift=0.1cm, inner sep=2.4pt, below=0.5cm of n,
		          name=o, shape=swig vsplit, swig vsplit={gap=4pt, inner line width right=0.3pt}]{
        					\nodepart{left}{$O$}
        					\nodepart{right}{$o$}}
		node[format, shape=ellipse, right =0.5cm of n] (m) {$M(n,o)$}
		node[format, shape=ellipse, right =0.5cm of m] (l) {$L(n,o)$}
		node[format, shape=ellipse, below=0.5cm of m] (y) {$Y(n,o)$}	
		node[below=0.3cm of y]{(c)}				
		(a) edge[red,ultra thick] (n)
		(a) edge[red, ultra thick] (o)
		(n) edge[blue] (m)
		(n) edge[blue,out=30,in=150] (l)
		(o) edge[blue,out=300,in=240,looseness=1] (l)
		(o) edge[blue] (y)
		(m) edge[blue] (y)
		(l) edge[blue] (m)
		(l) edge[blue] (y)
			;
\end{scope}
\end{tikzpicture}
\end{center}
\caption{Three different population SWIGs associated with the three expanded DAGs shown in Figure \protect\ref{fig:anomlpath}.
Under the SWIGs (a), (b) the effects of $N$ and $O$ on $Y$ are separable so that the distribution of $Y(n,o)$ in the four arms  are identified given the distribution of $Y(n\!=\!x,o\!=\!x)$ although the identification formulae differ;
(c) A SWIG under which the four arms $Y(n,o)$ are not identified from the two arms $Y(n\!=\!x,o\!=\!x)$.
 \label{fig:ymlno-swig}}
\end{figure}
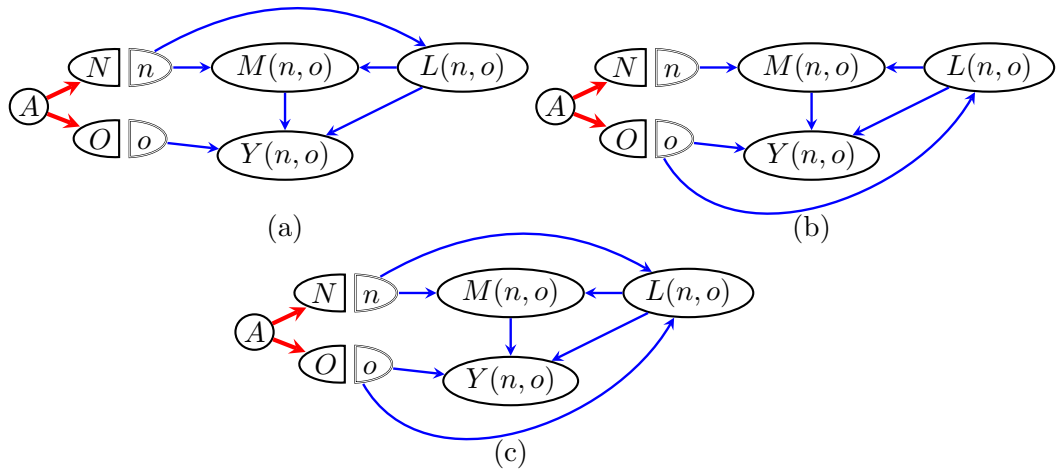

\subsection{Expanded Graphs For A Single Treatment}\label{sec:expanded-single}


We formally define expanded graphs as follows:\par
Given a DAG $\G$ with a single treatment variable $A$, an {\it expanded graph $\G^{ex}$ for $A$} is a DAG constructed by first adding a set of new
variables  $\{A^{(1)},\ldots , A^{(p)}\}$ corresponding to
a decomposition of the treatment $A$ into $p$ separate components 
(proposed by the investigator);
every variable $A^{(i)}$ is a child of $A$ with the same state space and 
$A^{(i)}(a) = a$, but $A$ has no other children in $\G^{ex}$; each child $C_j$ of $A$ in $\G$ has in $\G^{ex}$ a subset of $\{A^{(1)},\ldots , A^{(p)}\}$ as its set of parents.

\begin{lem}\label{lem:basic} Assume the distribution of the variables on an unexpanded DAG $\G$ is positive.
Under an FFRCISTG corresponding to an expanded (population) graph $\G^{ex}$ for treatment $A$, the intervention distribution 
$p(V(a^{(1)}=x^{(1)},\ldots,a^{(p)}=x^{(p)}))$ is identified by the g-formula applied to $\G^{ex}$ from the data on $\G$ if for every child $C_j$ of  $A$ in $\G$, the set of parents of $C_j$ in $\G^{ex}$ that are components of $A$ take the same value.\footnote{More formally, we require that for all $C_j \in \ch_{\G}(A)$, if $A^{(k)}, A^{(l)} \in \pa_{\G^{ex}}(C_j) \cap\{A^{(1)},\ldots , A^{(p)}\}$ then $x^{(k)} = x^{(l)}$.}
\end{lem}

\noindent{\it Proof:} As in the proof of Proposition \ref{prop:2-to-4-general}, consider the g-formula for $p(V(a^{(1)}=x^{(1)},\ldots,a^{(p)}=x^{(p)}))$
of Proposition \ref{prop:swig-pva-identified} in \citep{rrs21volume_id_arxiv} applied to the graph $\G^{ex}$.
The g-formula will be a function of the joint distribution of the observables since each term in the g-formula always conditions on a single value of $A$.
\medskip

\cite[Section 6.2]{robins10alternative} also described the special case in which each child $C_j$ of $A$ in $\G$ has exactly one component $A^{(j)}$ as a parent.\footnote{More formally, we have that for 
all children $C_j$ of $A$, 
$\pa_{\G^{ex}}(C_j) \cap \{A^{(1)},\ldots , A^{(p)}\} = \{A^{(j)}\}$.} We will refer to this as the {\it edge expanded graph for $A$}, which we often denote as $\G^{edge}$. 
In this case $p=|\ch_{\G}(A)|$ and in this special case $\G^{ex}$ corresponds to the graph formed from $\G$ by replacing each edge $A\rightarrow C_j$ with $A {\color{red}{\rightarrow}} A^{(j)} \rightarrow C_j$. Note that ${\G}^{edge}$ is unique.%
\footnote{Since the expanded graph ${\G}^{edge}$ for $A$ postulates a separate component of treatment corresponding to each child of $A$, such a graph will be unlikely to represent the substantive understanding of an investigator when $p$ is large.\label{foot:p-large}}
See Figure~\ref{fig:edge-expanded graph} for an example.

\begin{cor}\label{cor:pi} Under the assumptions of Lemma \ref{lem:basic}, if $\G^{ex}$ is the edge expanded graph $\G^{edge}$ for $A$
then for all assignments $x^{(1)},\ldots ,x^{(p)}$ $p(V(a^{(1)}=x^{(1)},\ldots,a^{(p)}=x^{(p)}))$ is identified
from the data on $\G$. 
\end{cor}

When the conditions of this Corollary hold, we will say, following the nomenclature in \citep{stensrud:separable-effects:2020}, that the treatment components  $\{A^{(1)},\ldots , A^{(p)}\}$ have {\it separable effects}. 

\subsection{On The Substantive Relationship Between Different $\G^{ex}$ Graphs And $\G^{edge}$}\label{subsec:gex-gedge}
In the context of the smoking cessation trial recall that the two expanded graphs in Figure \ref{fig:anomlpath} (a),(b) led to different identifying formulae for $P(Y(n=x,o=x^*))$ given, respectively, by Equations
(\ref{eq:swig-id-result-a}) and (\ref{eq:swig-id-result-b}). The identifying formulae also arise in the context of the graph $\G^{edge}$ shown in Figure \ref{fig:edge-expanded graph}. Specifically, $p(Y(a^{(1)}=x,a^{(2)}=x, a^{(3)}=x^*))$ and $p(Y(a^{(1)}=x^*,a^{(2)}=x, a^{(3)}=x^*))$ are identified by Equations (\ref{eq:swig-id-result-a}) and (\ref{eq:swig-id-result-b}) respectively.
The FFRCISTG models associated with the expanded graphs shown in Figure{s}~\ref{fig:anomlpath}~(a) and (b) correspond to distinct mutually exclusive causal structures, which lead to different identifying formulae for the distribution of the counterfactual $Y(n=x,o=x^*)$ in an arm of the four arm $(N,O)$ trial in which one intervenes to set $n=x$, and $o=x^*$.
However, given the FFRCISTG model corresponding to the graph in Figure \ref{fig:edge-expanded graph}, we are able to interpret the 
identifying expressions (\ref{eq:swig-id-result-a}) and (\ref{eq:swig-id-result-b})
 as identifying two {\it different} interventions on $A^{(1)}$, $A^{(2)}$, and $A^{(3)}$ on a single graph $\G^{edge}$. 
\par
\cite{robins10alternative} note that the above may seem to, but do not, contradict one another.
 Recall that $N$ and $O$ represent the substantive variables recording the presence or absence of nicotine and all other cigarette components.
 \medskip
  
 Suppose we further divide the other cigarette components into {t}ar ($T$) and cigarette components other than tar and nicotine ($O^*$). Thus substantively setting $O$ to a value corresponds to setting both $O^*$ and $T$ to that value. Furthermore,
 the graph $\G^{edge}$ in Figure~\ref{fig:edge-expanded graph} being an FFRCISTG implies that the graph $\G^{ex}$ in Figure~\ref{fig:anomlpath}~(b) formed by (re)combining $O^*$ and $T$ is an FFRCISTG. Thus an intervention setting $(O=x,N=x^*)$ on Figure~\ref{fig:anomlpath}~(b) substantively corresponds to the intervention  on the graph $\G^{edge}$ in 
Figure~\ref{fig:edge-expanded graph} with $O^*=A^{(3)}=x$, $T=A^{(1)}=x$, $N=A^{(2)}=x^*$. Thus these interventions in this $\G^{ex}$ and in $\G^{edge}$ give the same identifying formula (\ref{eq:swig-id-result-b}).
\par
Given {that} we have locked in the substantive interpretation of the $A^{(j)}$ in Figure \ref{fig:edge-expanded graph}, the intervention $O^*\equiv A^{(3)}=x$, $T\equiv A^{(1)}=x^*$, $N\equiv A^{(2)}=x^*$ on Figure \ref{fig:edge-expanded graph} corresponds to the intervention in which $N$ and $T$ are set to the same value and thus does not represent a joint intervention on the substantive variables $N$ and $O$ (since $O = O^* = T$ as random variables in the observed data {distribution}).
However, the identifying formula for this  intervention{,} if Figure \ref{fig:edge-expanded graph} were the causal graph{,} happens to have the same identifying formula (\ref{eq:swig-id-result-a}) as the intervention setting $O=x$, $N=x^*${,} if Figure \ref{fig:anomlpath}(a) were the causal graph. 
This might seem surprising since, under the substantive meanings of the components $(A^{(1)},A^{(2)},A^{(3)})$, specifically, $A^{(2)}\equiv N$, Figure  \ref{fig:edge-expanded graph} is compatible with Figure \ref{fig:anomlpath}(b) and {\it not} Figure \ref{fig:anomlpath}(a). However, it is {\it not} surprising from a formal perspective, if we notice that by considering a DAG with the same {\it structure} as Figure \ref{fig:anomlpath}(a), but in which $N$ is replaced
by a variable $N^\dag\equiv N\!\times\! T$ indicating the presence of both {t}ar {\it and} {n}icotine, then we may represent the intervention that sets $O^*=x$, $N=T=x^*$ via an intervention on $N^\dag$ and $O^*$.\footnote{Though the graph constructed in this way has the same topology as Figure \ref{fig:anomlpath}(a), it represents a different substantive hypothesis, since a component $T$ of $N^\dag$ is a parent of $L$ and not $N$ itself.}


\medskip

If, instead of dividing $O$, one divides {n}icotine {(}$N${)} into sub-components corresponding to two different isotopes
and re-defines the variables  in Figure~\ref{fig:edge-expanded graph} as $A^{(1)}\equiv$ {\it {n}icotine {i}sotope} 1, $A^{(2)}\equiv$ {\it {n}icotine {i}sotope} 2, $A^{(3)}\equiv O$ then the mirror image of the above holds. Specifically, Figure~\ref{fig:edge-expanded graph} is compatible with Figure~\ref{fig:anomlpath}~(a),\footnote{%
 Figure~\ref{fig:anomlpath}~(a) with the variables $N$ and $O$ 
 (not $N^\dag$ and $O$).} and not Figure~\ref{fig:anomlpath}~(b).

Note however, that there is no way to re-define $A^{(1)}$, $A^{(2)}$, and $A^{(3)}$ such that Figure~\ref{fig:edge-expanded graph} is compatible with 
 Figure~\ref{fig:anomlpath}~(c). Specifically, an intervention setting $N=x$ and $O=x^*\neq x$ cannot be represented via an intervention on 
$(A^{(1)}, A^{(2)}, A^{(3)})$, since in Figure~\ref{fig:anomlpath}~(c) $L$ has two parents $N$ and $O$ while in 
Figure~\ref{fig:anomlpath}~(c) $L$  has only one. Of course this must be the case because the intervention on $N$ and $O$ in Figure~\ref{fig:anomlpath}~(c) is not identified from the observed data.  In contrast, by Corollary \ref{cor:pi} any intervention on $A^{(1)},A^{(2)},A^{(3)}$ on graph Figure~\ref{fig:edge-expanded graph} is identified. 

\par
Lastly, note that $\G^{edge}$ assumes that $A^{(1)}$, $A^{(2)}$, and $A^{(3)}$ each directly affect only $L$, $M$ and $Y$, respectively;
this hypothesis could, in principle, be tested if one were to perform an eight arm $(A^{(1)},A^{(2)},A^{(3)})$  trial.\\
%

In general, if  $\G^{ex}$ is an FFRCISTG with separable (hence, identified) effects  and further $\G^{edge}$ is an FFRCISTG, then
the counterfactual variables in $\G^{ex}$  may be obtained from those in $\G^{edge}$ by imposing the equality $a^{(i)} = a^{(j)}$ whenever the corresponding children $C_i$ and $C_j$ of $A$ in $\G$ share a common parent in $\G^{ex}$. Note that this is directly analogous to the way in which the counterfactual variables $V_i(a)$ in  $\G(a)$ are equal to the counterfactuals $V_i(n=a,o=a)$ present in $\G(n,o)$.


\subsection{Generalizations}\label{subsec:generalization3}

The foregoing development may be further generalized in several ways:
\begin{itemize}
\item[(a)] Rather than having data from a single treatment variable $A$,  we may consider a study in 
which there were multiple treatment variables 
$\{A_1,\ldots ,A_k\}$. In this setting it may be of interest to attempt to identify the distribution $V(n_1,o_1,\ldots ,n_k,o_k)$ of a hypothetical future study 
where $N_i$ and $O_i$ are components of $A_i$ such that in the original study $A_i=O_i=N_i$. 
See Figure~\ref{fig:more-general}. More generally each $A_i$ may have $p_i$ components. The natural generalization of Lemma \ref{lem:basic} holds.%
\footnote{That is, we have identification if, for each $i$ and each child $C$ of $A_i$ in $\G$,  the subset of the $p_i$ components of $A_i$ that are parents
of $C$ in the expanded graph $\G^{ex}$ take the same value.}
\item[(b)] We may consider a setting in which some variables ($H$) in the underlying causal DAG $\G(V\cup H)$ are not observed; though variables we intervene on are observed, so $A\subseteq V$.



Here, we proceed in two steps.  In the first step, we check identification of a standard interventional distribution.  Specifically, we construct an ADMG $\G^{ex}$ containing the variables $\{N_1,O_1,\ldots ,N_k,O_k\}$, such that $N_i$ and $O_i$ have only  $A_i$ as a parent; the only edges with an arrowhead at $N_i$ and $O_i$ are of the form $A_i {\color{red}{\rightarrow}} $ while the only edges out are of the form $\rightarrow C$ where $C$ is a child of $A$ in $\G$.
  We then apply the extended ID algorithm described in \citep{rrs21volume_id_arxiv} to first determine whether {$p(V(n_1,o_1,\ldots ,n_k,o_k))$} 
would be identified given a positive distribution  $p(V \cup \{N_1,O_1,\ldots ,N_k,O_k\})$ over the observed variables and the treatment components.\footnote{This would correspond to an observational study where the variables
$V \cup \{N_1,O_1,\ldots ,N_k,O_k\}$ are observed in a population in which 
$N_i$ and $O_i$ are no longer deterministic functions of $A_i$, but for which all other one-step ahead counterfactuals remain the same; variables in $H$ are not observed.}

In the second step, we check whether the identification would hold under the weaker conditions where we only have access to a positive distribution on $p(V)$. This corresponds to
(i) making sure that identification of every term in the identifying formula given by the ID Algorithm via the inductive application of Proposition 41.5 from 
the companion paper \citep{rrs21volume_id_arxiv} ensures that the splitting operation is applied to any $A_i \in A$ before any $N_i$ or $O_i$ (this ensures that the positivity requirement for Proposition 41.5 is met), and (ii) confirming that for every $(N_i,O_i)$, in the identifying formula given by the ID Algorithm\footnote{See Equation (\ref{eqn:line-4}) and subsequent discussion in \citep{rrs21volume_id_arxiv}.}
 there is no district $D$ which has both $N_i$ and $O_i$ as parents; \cite{shpitser13cogsci} terms such districts, which violate this condition, {\it recanting districts}.\footnote{Note that in the special case where $D=\{V_i\}$ is a singleton, then $D$ will be a recanting district
in $\G$ if and only if $N_i$ and $O_i$ have a common child in $\G^{ex}$. In this case $V_i$ is a ``recanting witness'' as defined in
\cite{chen05ijcai}; see also Section~\protect\ref{sec:path-specific}. 
Thus, when no hidden variables exist, the first step in (b) always succeeds; hence,
as implied Lemma \ref{lem:basic}, identification fails if and only if there exists an $N_i$ and $O_i$ that have a common child.}

A simple example of such a structure arises when there is an unobserved common cause of $M$ and $Y$, as shown in Figure \ref{fig:amyl}(b): though the four distributions $E[Y(n,o)]$ are identified given a four arm $(N,O)$ trial, the distributions for which $n\neq o$ are not identified solely given data on $(A,M,Y)$. This is because 
$M\leftarrow H \rightarrow Y$ forms a district and both $N$ and $O$ are parents of this district, hence it is recanting.

If the two step procedure yields identification, the resulting functional structurally resembles the functional obtained from the ID algorithm, except each term in the functional that depends on treatments is evaluated at its own treatment value, corresponding to either $n_i$ or $o_i$ (but never both at once).
\end{itemize}
See Section~\ref{sec:path-specific} below for a general method of addressing all complications above simultaneously.


%
%
%


\begin{figure}
\begin{center}
		\begin{tikzpicture}[>=stealth, node distance=1.2cm]
		\tikzstyle{format} = [draw, thick, circle, minimum size=4.0mm,
		inner sep=1pt]
		\tikzstyle{unode} = [draw, thick, circle, minimum size=1.0mm,
		inner sep=0pt,outer sep=0.9pt]
		\tikzstyle{square} = [draw, very thick, rectangle, minimum size=4mm]
		\begin{scope}
		\path[->,  line width=0.9pt]
		node[format,  inner sep=1.8pt, shape=ellipse] (a) {$A$}
		node[format, xshift=0cm, yshift=0.1cm, inner sep=2.4pt, right of = a,
		          name=n, shape=swig vsplit, swig vsplit={gap=4pt, inner line width right=0.3pt}]{
        					\nodepart{left}{$N$}
        					\nodepart{right}{$n$}}
		node[format, xshift=0cm, yshift=0.1cm, inner sep=2.4pt, below=0.5cm of n,
		          name=o, shape=swig vsplit, swig vsplit={gap=4pt, inner line width right=0.3pt}]{
        					\nodepart{left}{$O$}
        					\nodepart{right}{$o$}}
		node[format, shape=ellipse, right =1cm of n] (m) {$M(n)$}
		node[format, shape=ellipse, below=1cm of m] (y) {$Y(n,o)$}			
		(a) edge[red,ultra thick] (n)
		(a) edge[red, ultra thick] (o)
		(n) edge[blue] (m)
		(o) edge[blue] (y)
		(m) edge[blue] (y)
		;
		\end{scope}
%
%
%
%
		\end{tikzpicture}
\end{center}
\caption{A SWIG derived from the expanded DAG $\G^{ex}$ in Figure \protect\ref{fig:amyno} (c), under the assumption that the counterfactual variables
obey the NPSEM associated with $G^{ex}$. Consequently, in contrast to Figure \protect\ref{fig:amyno-swig}(b), the graph contains $M(n)$
rather than $M(n,o)$.}\label{fig:amyno-swig-individual}
\end{figure}
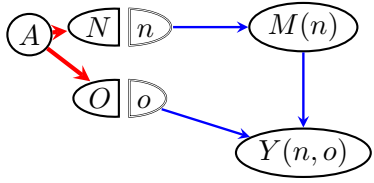

\subsection{Identification Of Cross-World Nested Counterfactuals Of DAG $\G$ Under An FFRCISTG Model For Its Expanded Graph $\G^{ex}$}
\label{sec:back-to-pde}

From Section \ref{subsec:separability} to this point, we only studied the distribution of counterfactuals associated with interventions on $(N_i,O_i)$ or, more generally,
$(A^{(1)},\ldots , A^{(p)})$; no other counterfactuals associated with an expanded graph $\G^{ex}$ were mentioned. As argued in Section \ref{subsec:separability}, for most purposes these counterfactuals
constitute an adequate basis for formulating contrasts relating to the mediation of effects.
However, since the prior literature on mediation has been formulated in terms of cross-world nested counterfactuals 
associated with the original (unexpanded) DAG $\G$, we return to our earlier discussion.

\subsubsection*{The PDE}
Recall that in Section \ref{sec:mediation} we related the PDE from the NPSEM associated with 
Figure~\ref{fig:amyno}~(a) to a four arm $(N,O)$ trial under the FFRCISTG
associated with the SWIG $\G^{ex}(n,o)$  shown in Figure~\ref{fig:amyno-swig}~(b).
This may be broken down into two steps:
\begin{itemize}
\item[(1)] Show that $E[Y(n=0,o=1)]$ was identified from data on $p(A,M,Y)$ via Equation (\ref{eq:two-to-four}).
This step only required that $\G^{ex}$ was a population FFRCISTG.\footnote{Thus this step does not require that the counterfactual variables follow an NPSEM associated with $\G^{ex}$.} That is, this identification follows from the deterministic relationship $N(a)=O(a)=a$ together with the population level conditions (\ref{eq:conditions-for-4-to-2a}) and (\ref{eq:conditions-for-4-to-2b}), without requiring an $M$ counterfactual be well-defined.
\item[(2)] Next show that $Y(a=1,M(a=0)) = Y(n=0,o=1)$ holds under the individual level no direct effect 
assumptions encoded in the NPSEM associated with $\G^{ex}$ in Figure~\ref{fig:amyno-swig-individual}. Note that this step
does not require that Figure~\ref{fig:amyno}~(c) is an FFRCISTG, only that it be an NPSEM associated with $\G^{ex}$.

In more detail, the NPSEM associated with $\G^{ex}$ implies the following:
\begin{itemize}
\item[(i)]  $M(n) = M(a,n,o)$;
\item[(ii)]  $Y(o,m) = Y(a,n,o,m)$.
\end{itemize}

Under conditions (i) and (ii) we have that:
\begin{equation}
\begin{array}{rcccl}
M(a=0) &= & M(N(a=0))&=&M(n=0),\\[6pt]
Y(a=1,m) & = &Y(O(a=1),m)& =& Y(o=1,m),\\[6pt]
Y(a=1,M(a=0)) &= &Y(o=1,M(n=0))&=&Y(n=0,o=1).
\end{array}
\end{equation}
Consequently, under the assumptions (i), (ii) it follows that:
\begin{equation}
P(Y(a=1,M(a=0))) = P(Y(n=0,o=1)).
\end{equation}
\end{itemize}

\noindent{\bf Remarks:} 
\begin{itemize}
\item[1.] If $N(a)=O(a)=a$ and conditions (i) and (ii) hold, then $PDE = E[Y(n=0,o=1)] - E[Y(n=0,o=0)]$ even if conditions 
(\ref{eq:conditions-for-4-to-2a}) and (\ref{eq:conditions-for-4-to-2b}) fail, for example due to $(M,Y)$ confounding.  This is the situation discussed in Section \ref{sec:mediation} where data from the four arm $(N,O)$ trial makes it possible to estimate the PDE and hence determine whether it equals the mediation formula.\footnote{Note, however, that if $P(Y(n=0,o=1))$ does not equal the first expression in the mediation formula it is possible that this is solely because $(N,O)$ do not satisfy (i) and (ii) but that there are other subcomponents of $A$, say $(N^*,O^*)$ that do satisfy (i) and (ii).}
\item[2.] The conditions (\ref{eq:conditions-for-4-to-2a}) and (\ref{eq:conditions-for-4-to-2b}) alone, i.e.~without (2) above,
 are not sufficient to identify the PDE{.}\footnote{
 This is because 
 \begin{align}
 E[Y(a=1,M(a=0))] &= E[Y(n=1,o=1, M(n=0,o=0))] \\
& = \sum_{m}   E[Y(n=1,o=1, m)\,|\, M(n=0,o=0)=m] p(M(n=0,o=0)=m),
\end{align}
but this latter conditional expectation term is cross-world in terms of the counterfactuals in $\G^{ex}$,
although, as noted, they do identify $E[Y(n=0,o=1)] - E[Y(n=0,o=0)]$.
}
\item[3.(i)] If condition (i) fails, so that $O$ has a (population) direct effect on $M$ (relative to $A$, $N$) then the counterfactual $M(n,o)$ is not a function of the 
original one-step-ahead counterfactuals $Y(a,m)$ and $M(a)$. This can be seen from the fact that whereas $M$ has a single parent $A$ in Figure~\ref{fig:amyno}~(a), under the elaboration that includes $N$ and $O$ it now has two: $\{N,O\}$.
\item[3.(ii)] Similarly, if condition (ii) fails, so that $N$ has a direct effect on $Y$ (relative to $A$, $M$, $O$) then the counterfactual $Y(n,o)$ is not a function of the original one-step-ahead counterfactuals $Y(a,m)$ and $M(a)$.  $Y$ has two parent $\{A,M\}$ in Figure~\ref{fig:amyno}~(a), under the elaboration that includes $N$ and $O$ it would have three $\{N,O,M\}$.
\end{itemize}

\subsubsection*{Example: The River Blindness Studies}
Returning to the River Blindness study, note that intervening to set $s=1$ and $a=0$ gives $Y(s=1,a=0) = Y(a=1,M(a=0))$ as random variables. Hence the PDE is identified from
data in a four arm $(A,S)$ trial.\footnote{Recall that in $Y(s=1,a=0)$, $A$ is serving as `$N$'; see Figure \ref{fig:amyno}(e) and Footnote {\protect\ref{foot:detectable}}.}
 It is not identified from data on $A$, $M$, $Y$ because $M\leftrightarrow Y$ forms a ``recanting district,'' as defined in \citep{shpitser13cogsci}.
Note that had the identification of the PDE failed due to a recanting witness, that is, $A$ and $S$ having a common child, then additional interventions on the variables in the graph would not have led to identification. Finally we note that, owing to the context specific conditional independences in this example, 
the distribution $p(Y(s\!=\!1,a\!=\!0)) = {p(Y(a = 0, M(a=1)))}$, which occurs in the {t}otal {d}irect {e}ffect is identified given data from the two arms $p(Y(s\!=\!x,a\!=\!x))$, $x\in \{0,1\}$ by the formula given in (\ref{eqn:determinism}), and hence also from $p(A,M,Y)$; see Appendix Section~\ref{subsec:pde-not-identified-river-blindness}.

\subsubsection*{Counterfactuals Related To
The DAG In Figure~\protect\ref{fig:amyl}~(a)}

Recall that $E[Y(n\!=\!0,o\!=\!1)]$ is identified under the FFRCISTG models associated with the graphs in Figures~\ref{fig:anomlpath}~(a) and (b), but not (c).

\begin{figure}
\begin{center}
\begin{tikzpicture}[>=stealth, node distance=1.2cm]
\tikzstyle{format} = [draw, thick, circle, minimum size=4.0mm,
		inner sep=2pt]
		\tikzstyle{unode} = [draw, thick, circle, minimum size=1.0mm,
		inner sep=2pt,outer sep=0.9pt]
		\tikzstyle{square} = [draw, very thick, rectangle, minimum size=4mm]
\begin{scope}
\path[->,  line width=0.9pt]
		node[format, shape=ellipse] (a) {$A$}
		node[format, shape=ellipse, right =0.5cm of a] (n) {$A^{(2)}$}
		node[format, shape=ellipse, below = 0.5cm of n] (o) {$A^{(3)}$}
		node[format, shape=ellipse, above = 0.5cm of n] (t) {$A^{(1)}$}
		node[format, shape=ellipse, right =2cm of a] (m) {$M$}
		node[format, shape=ellipse, right =1cm of m] (l) {$L$}
		node[format, shape=ellipse, below=1cm of m] (y) {$Y$}	
		(a) edge[red,ultra thick] (n)
		(a) edge[red, ultra thick] (o)
		(a) edge[red, ultra thick] (t)
		(n) edge[blue] (m)
		(t) edge[blue,out=30,in=150] (l)
		(o) edge[blue] (y)
		(m) edge[blue] (y)
		(l) edge[blue] (m)
		(l) edge[blue] (y)
			;
\end{scope}
\end{tikzpicture}
\end{center}
\caption{The edge expanded graph $\G^{edge}$ associated with the DAG shown in Figure \protect\ref{fig:amyl} (a).}
\label{fig:edge-expanded graph}
\end{figure}
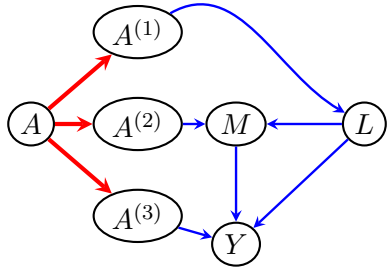

Let $Y(a,l,m)$, $M(a,l)$ and $L(a)$ denote the one-step-ahead
counterfactuals associated with the graph in Figure~\ref{fig:amyl}~(a).
It follows from the deterministic counterfactual relation $N\left(
a\right) =O\left( a\right)=a$ and the NPSEM associated with Fig.~\ref{fig:amyl}~(a), and its associated expanded graph
${\cal G}^{\text{ex}}$ in Figure~\ref{fig:anomlpath}~(a)
that the random variable 
\begin{equation*}
Y(n\!=\!0,o\!=\!1) = Y(o=1,L(n=0),M(n=0,L(n=0)))
\end{equation*}%
associated with the NPSEM in Figure~\ref{fig:anomlpath}~(a) can be written
in terms of the counterfactuals associated with the graph in Figure~\ref%
{fig:amyl}~(a) as the cross-world counterfactual
\begin{equation}\label{eq:crossworld-pde-27}
Y(a\!=\!1,L(a\!=\!0),M(a\!=\!0)) = Y(a\!=\!1,L(a\!=\!0),M(a\!=\!0,L(a\!=\!0))).
\end{equation}
Similarly, if we assume that $\G^{edge}$ of Figure~\ref{fig:edge-expanded graph} is an NPSEM, the counterfactual (\ref{eq:crossworld-pde-27})  also equals, as a random variable, the counterfactual $Y(A^{(1)}=0,A^{(2)}=0,A^{(3)}=1)$.
Thus if either Figure~\ref{fig:anomlpath}~(a) or Figure~\ref{fig:edge-expanded graph} represented the FFRCISTG generating the data, the distribution of Equation
(\ref{eq:crossworld-pde-27}) is identified by the same formula (\ref{eq:swig-id-result-a}).

Likewise, 
\begin{equation*}
Y(n\!=\!0,o\!=\!1) = Y(o=1,L(o=1),M(n=0,L(o=1)))
\end{equation*}%
associated with the graph in Figure~\ref{fig:anomlpath}~(b) equals (as a random variable) the cross-world counterfactual
\begin{equation}\label{eq:crossworld-pde-28}
Y(a=1,L(a=1),M(a=0,L( a=1) ))
\end{equation}%
associated with the graph in Figure~\ref%
{fig:amyl}~(a).  Again, if we assume that $\G^{edge}$ of Figure~\ref{fig:edge-expanded graph} is an NPSEM, the counterfactual (\ref{eq:crossworld-pde-28})  also equals $Y(A^{(1)}=1,A^{(2)}=0,A^{(3)}=1)$ (as a random variable).
Thus in this case if either Figure~\ref{fig:anomlpath}~(b) or Figure~\ref{fig:edge-expanded graph} represented the FFRCISTG generating the data, the distribution of 
(\ref{eq:crossworld-pde-28}) is identified by the same formula (\ref{eq:swig-id-result-b}).

 In contrast $E[Y(n\!=\!0,o\!=\!1)] $ associated
with the graph in Figure~\ref{fig:anomlpath}~(c) is not the mean of any
counterfactual defined from $Y(a,l,m)$, $M(a,l)$, and $L(a)$ under the graph
in Figure~\ref{fig:amyl}~(a) since $L$, after intervening to set $n=0$, $o=1,$
is neither $L(a=1)$ nor $L(a=0)$ as both imply a counterfactual for 
$L$ under which $n=o$.

Furthermore, the parameter occurring in the PDE
\begin{equation*}
E\left[ Y(a=1,M(a=0))\right] =E\left[ Y(a=1,L(a=1),M(a=0,L(a=0)))\right] 
\end{equation*}
and associated with the graph in Figure \ref{fig:amyl}(a) is not identified
under the FFRCISTGs 
associated with any of the three graphs in
Figures~\ref{fig:anomlpath}~(a), (b) and (c). 
This is because all three of these graphs are compatible with the NPSEM in Figure~\ref{fig:amyl}~(a)
under which, by recursive substitution, 
we have the following equality $Y(a_1,M(a_0)) = Y(a_1,L(a_1),M(a_0,L(a_0)))$, as random variables.
But the counterfactual $Y(a_1,L(a_1),M(a_0,L(a_0)))$, since it involves $L(a_0)$ {\it and} $L(a_1)$, does {\it not} correspond to an intervention on $n$ and $o$
under any of the expanded graphs in Figures~\ref{fig:anomlpath}~(a), (b) or (c).
Similarly, $Y(a_0,M(a_1))$ is not identified under any of these graphs. 
These results follow from the fact that $L$ is a recanting witness for $A$ in Figure~\ref{fig:amyl}~(a).

%
%
%
%
%
%

In the following section, we define path specific effects associated with the NPSEM model. We show that the two identified nested cross-world counterfactuals are identified path-specific effects
under the NPSEM-IE for the graph in Figure~\ref{fig:amyl}~(a).  In particular,
$E[Y(a=1,L(a=0),M(a=0))]$ is the path-specific effect associated with the path $A \to Y$ 
and
$E[Y(a=1,L(a=1),M(a=0,L(a=1)))]$ is the path-specific effect associated with the paths $A \to Y$, $A \to L \to Y$, and $A \to L \to M \to Y$.
Thus, for Pearl, identification of these nested cross-world counterfactuals associated with  Figure~\ref{fig:amyl}~(a)
 follows from the assumption that the distribution of the counterfactuals obeys the NPSEM-IE model associated with the graph in Figure~\ref{fig:amyl}~(a).
In contrast,
the identification of these cross-world counterfactuals under an FFRCISTG model follows from two different extensions of Pearl's original story: the first is identified under the extension in which $N$ but not $O$ is a cause of $L$ and the second from the extension that $O$ but not $N$ is a cause of $L$.  The respective identifying formulae are the same under both theories.

\section{Path-Specific Counterfactuals}\label{sec:path-specific}

In Section 
\ref{sec:mediation}
we considered different notions of {\it direct} effect, which led to the notion of nested counterfactuals and the PDE, which is identified under the NPSEM-IE associated with the graph in
Figure~\ref{fig:amyno}~(a) via the associated {\emph{mediation formula}}. 
In Section \ref{subsec:separability}, following \cite{robins10alternative}, we discuss the notion of separability of effects in the sense of \cite{stensrud:separable-effects:2020} via an expanded $(N,O)$ graph. We showed that the counterfactuals defining the PDE were equal to ordinary (non cross-world) interventional counterfactuals in the NPSEM given by the expanded graph (with determinism). We also showed that under the corresponding FFRCISTG these effects were identified by the mediation formula. In this section we now consider path specific effects which generalize the notion of direct and indirect effects.

In the simplest setting, the intuition behind an indirect effect is to consider all paths from $A$ to $Y$ {\it other than} the edge $A\rightarrow Y$. This can be generalized to settings where the effect along a particular {\it set} of causal paths from $A$ to $Y$ is of interest.
In what follows we will show that each such path specific effect will correspond to a cross-world counterfactual contrast associated with the (original) graph $\G$. 
We will see that in the absence of recanting witnesses, these cross-world counterfactuals are equal (as random variables) to interventional counterfactuals in 
the NPSEM associated with $\G^{edge}$, the edge expanded graph associated with the set of treatment variables $A$.\footnote{As noted earlier,
  the counterfactuals associated with  $\G^{edge}$ may not have clear substantive meaning. For $\G^{edge}$ to be substantive
  it is necessary for each treatment variable $A$ to be decomposable into components each of which could, in principle, be intervened on (separately) and affect one and only one of its children; see Footnote~{\protect\ref{foot:p-large}}, Section \protect\ref{subsec:gex-gedge} for further discussion. 
  The graph $\G^{edge}$ may still be useful as a formal construction.}
Consequently, these path specific 
counterfactuals will be identified if and only if the corresponding intervention is identified under the FFRCISTG associated with 
  $\G^{edge}$. Thus all identifying formulae for path-specific cross-world counterfactuals on $\G$ may be derived from $\G^{edge}$.
Further, it follows that all identifying formulae for path-specific cross-world counterfactuals may also be obtained under the assumption that $\G$ is an NPSEM-IE.



The general theory developed by \cite{shpitser13cogsci} associates a random variable with each subset of causal paths between a treatment $A$ and an outcome $Y$.
The intuition is that this subset of proper\footnote{A proper causal path intersects the set $A$ only once at the source node.} causal paths from $A$ to $Y$ denoted $\pi$ remain active, while all other causal paths, denoted $\overline{\pi}$, from $A$ and $Y$ are blocked.\footnote{It is important to understand that the predicates ``active'' and ``blocked'' are applied to paths. In particular, it is possible for every edge and vertex on a blocked path to also be present on some active path, and vice-versa.}
Next, pick a pair of value sets $a$ and $a'$ for elements in $A$; $a$ will be associated with active paths, $a'$ with those that are blocked.

For any $V_i \in V$, define the potential outcome $V_i(\pi,a,a')$ by setting $A$ to $a$ for the purposes of \st{paths in $\pi$ that end in $V_i$}
{proper causal paths from $A$ to $V_i$ in $\pi$}, and setting $A$ to $a'$ for the purposes of proper causal paths from $A$ to $V_i$ not in $\pi$.\footnote{Note that it follows from the definition of proper causal path that each path in $\pi$  is associated with a unique vertex in $A$; similarly for each path in $\overline{\pi}$. (Two paths may be associated with the same vertex in $A$.)}
  Formally, the definition is as follows, for any $V_i \in V$:
\begin{align}
\label{eqn:pse}
V_i(\pi, a, a') &\equiv a \;\text{ if }\; V_i \in A,\\
V_i(\pi, a, a') &\equiv 
V_i( \{ V_j(\pi, a, a') \mid V_j \in \pa^{\pi}_i \}, \{ V_j(a') \mid V_j \in \pa^{\overline{\pi}}_i \} )
\notag
\end{align}
where $V_j(a') \equiv a'$ if $V_j \in A$, and given by recursive substitution otherwise,  
$\pa^{\pi}_i$ is the set of parents of $V_i$ along an edge which is a part of a path in $\pi$, and
$\pa^{\overline{\pi}}_i$ is the set of all other parents of $V_i$.

A counterfactual $V_i(\pi, a, a')$ is said to be \emph{edge inconsistent} if {for some edge $A_k\rightarrow V_j$ in $\cal G$}, counterfactuals of the form
$V_j(a_k, \ldots)$ and $V_j(a_k', \ldots)$ occur in $V_i(\pi, a, a')$, otherwise it is said to be \emph{edge consistent}. In the former case $V_j$ is said to be {\it a recanting witness (for $\pi$)}. It is simple to verify using Equation (\ref{eqn:pse}) that edge consistent counterfactuals are precisely those where no paths in $\pi$ and $\bar{\pi}$ share the initial edge. \cite{shpitser13cogsci,shpitser15hierarchy}
have shown that a joint distribution $p(V(\pi, a, a'))$ containing an edge-inconsistent counterfactual $V_i(\pi, a, a')$ is not identified in 
the NPSEM-IE (nor weaker causal models) in the presence of a recanting witness. 

As an example, consider the graph shown in Figure~\ref{fig:amyl}~(a) and the 
counterfactual given in Equation (\ref{eq:crossworld-pde-27}) which corresponds to the path $\pi_1 = \{ A \to Y \}$:
\[
Y(\pi_1, a\!=\!1, a\!=\!0) \equiv Y(a=1,L(a=0),M(a=0, L(a=0))).
\]
The counterfactual associated with the paths $\pi_2 = \{ A \to Y, A \to L \to Y \}$ is given by:
\[
Y(\pi_2, a\!=\!1, a\!=\!0) \equiv Y(a=1,L(a=1),M(a=0, L(a=0))).
\]
Note that $Y(\pi_1,a\!=\!1,a\!=\!0)$ is edge consistent, while $Y(\pi_2,{a\!=} 1,{a\!=} 0)$ is edge-inconsistent due to the presence of $L(a=0)$ and $L(a=1)$.\footnote{The above development may be generalized to $k$ different assignments rather than two, by partitioning the set of paths into $k$. These and other generalizations are termed \emph{path interventions} by \cite{shpitser15hierarchy}.} 
%
 
This result is proved in \citep{shpitser15hierarchy}:
\begin{thm}\label{thm:edge-con-nspem-ie}
	If $V(\pi, a, a')$ is edge consistent, then under the NPSEM-IE for the DAG $\mathcal{G}$,
	\begin{align}
	p(V(\pi, a, a')) = \prod_{i=1}^K p(V_i \mid a \cap \pa^{\pi}_i, a' \cap \pa_i^{\overline{\pi}}, \pa^{\cal G}_i \setminus A).
	\label{eqn:edge-g}
	\end{align}
\end{thm}
As an example of such an identification consider the distribution $p(Y(\pi, a, a'))$ of the edge consistent counterfactual in Figure~\ref{fig:amyl}~(a). It follows from Theorem \ref{thm:edge-con-nspem-ie} that
 \begin{align*}
p(Y(\pi_1, a\!=\!1, a\!=\!0))&= p(Y(a=1,L(a=0),M(a=0, L(a=0))))\\
&= \sum_{m,l} p(Y\,|\,m, l, a=1) p(m\,|\, l,a=0)p(l\,|\,a=0),
 \end{align*}
a marginal distribution derived from (\ref{eqn:edge-g}).


In the following, we exploit an equivalence between 
edge consistent counterfactuals $V_i(\pi,a,a')$ and standard potential outcomes based on edge expanded graphs $\mathcal{G}^{{edge}}$,
already defined in the case of a single treatment variable in Section~\ref{sec:expanded-single}.\footnote{A similar construction was called the ``extended graph'' in \citep{malinsky19po}.} 
 In this section, we will abbreviate ${\cal G}^{{edge}}$ as ${\cal G}^e$ for conciseness.  The edge expanded graph both simplifies complex nested potential outcome expressions and enables us to leverage the prior result in \citep{shpitser06idc} to identify conditional path-specific effects.

We now extend the definition of expanded graph to sets of treatments $|A|>1$: Given an ADMG ${\cal G}(V)$, define for each $A_i \in A \subseteq V$ {a new} set of variables, 
$A_i^{\Ch} \equiv \{ A_i^{j} \mid V_j \in \Ch_i \}$ such that $\mathfrak{X}_{A_i^j} \equiv \mathfrak{X}_{A_i}$; thus for each directed edge $A_i \to V_j$ in ${\cal G}(V)$ from a treatment variable to its child, we have created a new variable $A_i^j$ with the same state space as $A_i$.
Denote the full set of new variables as $A^{\Ch} \equiv \bigcup_{A_i \in A} A_i^{\Ch}$.  We define the edge expanded
graph of ${\cal G}(V)$, written ${\cal G}^e(V\cup A^{\Ch})$, as the graph with the vertex set $V \cup A^{\Ch}$; the edge expanded graph contains all the edges in ${\cal G}$ except for the edges $A_i \rightarrow V_j$ that join a treatment variable $A_i$ to its child $V_j$, in addition we add the edges  $A_i \to A_i^j \to V_j$ (which thus ``replace'' the removed edge $A_i \rightarrow V_j$).

\begin{figure}
\begin{center}
		\begin{tikzpicture}[>=stealth, node distance=1.2cm]
		\tikzstyle{format} = [draw, thick, circle, minimum size=4.0mm,
		inner sep=1pt]
		\tikzstyle{unode} = [draw, thick, circle, minimum size=1.0mm,
		inner sep=0pt,outer sep=0.9pt]
		\tikzstyle{square} = [draw, very thick, rectangle, minimum size=4mm]
		\begin{scope}
		\path[->,  line width=0.9pt]
		node[format, shape=ellipse] (a) {$A_1$}
		node[format, shape=ellipse, right =0.5cm of a] (n) {$A_1^{(2)}$}
		node[format, shape=ellipse, below = 0.5cm of n] (o) {$A_1^{(3)}$}
		node[format, shape=ellipse, above = 0.2cm of n] (q) {$A_1^{(2)}$}
		node[format, shape=ellipse, right =2cm of a] (m) {$M$}
		node[format, shape=ellipse, below=1.5cm of m] (y) {$Y$}			
		node[format, shape=ellipse, right =2cm of m] (l) {$A_2$}
		node[format, shape=ellipse, left =0.5cm of l] (r) {$A_2^{(1)}$}
		node[format, shape=ellipse, below =0.2cm of r] (s) {$A_2^{(2)}$}

		(q) edge[blue,out=30,in=90] (l)
		(r) edge[blue] (m)
		(s) edge[blue] (y)

		(a) edge[red,ultra thick] (n)
		(a) edge[red, ultra thick] (o)
		(a) edge[red, ultra thick] (q)

		(l) edge[red, ultra thick] (r)
		(l) edge[red, ultra thick] (s)

		(n) edge[blue] (m)
		(o) edge[blue] (y)
		(m) edge[blue] (y)
		;
		\end{scope}
		\end{tikzpicture}
\end{center}
\caption{
 An edge expanded graph that considers components of the treatment variables $A_1$ and $A_2$.
}
\label{fig:more-general}
\end{figure}
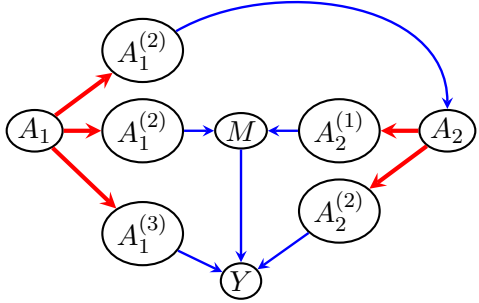

As an example, the edge expanded graph for the DAG in Figure~\ref{fig:amyno}~(a), with $A_1 = V$, is shown in Figure~\ref{fig:more-general}.  For conciseness, we will generally drop explicit references to vertices $V \cup A^{\Ch}$, and denote edge expanded graph of ${\cal G}(V)$ by ${\cal G}^e$. 

More generally, we associate a causal model with $\mathcal{G}^e$ as follows. 
Recall that we associated a set of one-step ahead potential outcomes ${\mathbb V}$ with the original graph $\mathcal{G}$. Similarly, we let ${\mathbb V}^e$ denote the set of  one-step ahead potential outcomes associated with  $\mathcal{G}^e$, constructed as follows: For every $V_i(\pa_i) \in {\mathbb V}$, we let $V_i(\pa^{e}_i)$ be in ${\mathbb V}^e$.  Note that this is well-defined, since $V_i$ in ${\cal G}$ and ${\cal G}^e$ share the number of parents, and the parent sets for every $V_i$ share state spaces.  In addition, for every $A_i^j \in A^{\Ch}$, we let $A_i^j(a_i)$ for $a_i \in {\mathfrak X}_{A_i}$ be in ${\mathbb V}^e$. \label{page:formal}

The edges $A_i \to A_i^j$ in $\mathcal{G}^e$ are understood to represent \emph{deterministic} {equality relationships, such that $A_i^j = A_i$.}
More precisely, every $A_i^j \in A^{\Ch}$ has a single parent $A_i$, and we let $A_i^j(a_i) = a_i$ so that $A_i^j(a_i)$ is a degenerate (constant) random variable
corresponding to a point-mass at $a_i$. 
The  FFRCISTG model associated with the edge expanded graph ${\cal G}^e$ includes these deterministic relationships.
Note that it follows that given a distribution $p(\mathbb{V})$ over the counterfactuals given by the NPSEM ${\cal G}$ there is a unique
distribution $p^e(\mathbb{V}^e)$ over the counterfactuals given by ${\cal G}^e$ that satisfies these deterministic relationships. 

%


We now show the following two results.  First, we show that an edge-consistent $V(\pi, a, a')$ may be represented without loss of generality by a counterfactual response to an intervention on a subset of $A^{\Ch}$ in ${\cal G}^e$ with the causal model defined above.  Second, we show that this response is identified by the same functional (\ref{eqn:edge-g}).

Given an edge consistent $V(\pi, a, a')$, define ${\cal G}^e$ for $A \subseteq V$.
We define $a^{\pi}$ that assigns $a_i$ to $A_i^j \in A^{\Ch}$ if $A_i \to V_j$ in ${\cal G}(V)$ is in $\pi$, and assigns $a_i'$ to $A_i^j \in A^{\Ch}$ if $A_i \to V_j$ in ${\cal G}(V)$ is not in $\pi$.  The resulting set of counterfactuals $V(a^{\pi})$ is well defined in the model for
$\mathbb{V}^e$, 
and we have the following result, proved in the Appendix. 

\begin{prop}
Fix a distribution $p(\mathbb{V})$ in the causal model for a DAG ${\cal G}(V)$, and consider the corresponding distribution $p^e(\mathbb{V}^e)$ in 
the FFRCISTG model
associated with a DAG ${\cal G}^e(V \cup A^{\Ch})$.  Then
for every $V_i \in V$, the random variable $V_i$ in the original model associated with ${\cal G}$ is equal to the random variable $V_i$ in the 
restrictd model associated with ${\cal G}^e$ that includes the equalities defining the variables in $A^{\Ch}$.
Moreover, for any edge consistent $\pi,a,a'$, the random variable $V_i(\pi,a,a')$ in the original model associated with ${\cal G}$ is equal to the random variable $V_i(a^{\pi})$ in the restricted model associated with ${\cal G}^e$.
\label{prop:equiv}
\end{prop}



\begin{thm}
	\label{thm:dag-ext-ffrcistg}
	Under the  FFRCISTG model associated with the edge expanded DAG ${\cal G}^e$,
	for any edge consistent $\pi,a,a'$:
	\begin{align}
	p^e(V(a^{\pi}))
	&= \prod_{i = 1}^K p^e(V_i \mid a^{\pi} \cap \pa^{{\cal G}^e}_i, \pa^{{\cal G}^e}_i \setminus A). \label{eq:yet-another-identifying-formula}
	\end{align}
\end{thm}
Note that since, by definition,  the distribution $p^e(V\cup {A}^{\Ch})$ is a deterministic function of $p(V)$, hence by Equation (\ref{eq:yet-another-identifying-formula})
$p^e(V(a^{\pi}))$ is identified by $p(V)$. Recall that if  ${\cal G}^e$  is an FFRCISTG model then this requires that all of the
interventions on the variables in $A^{\Ch}$ are well-defined.

\begin{cor}
	\label{cor:dag-ext}
	Under the NPSEM-IE model associated with the DAG ${\cal G}$, for any edge consistent $\pi,a,a'$:
	\begin{align*}
	p(V(\pi, a, a'))& = p^e(V(a^{\pi}))
	\;=\; \prod_{i = 1}^K p^e(V_i \mid a^{\pi} \cap \pa^{{\cal G}^e}_i, \pa^{{\cal G}^e}_i \setminus A), 
	\end{align*}
	where ${\cal G}^e$ is the edge expanded graph corresponding to $\G$.
\end{cor}
In fact, Corollary \ref{cor:dag-ext} provides an alternative proof of Theorem \ref{thm:edge-con-nspem-ie} since, by definition, for any edge consistent $\pi,a,a'$: 
\[
p(V_i \mid a \cap \pa^{\pi}_i, a' \cap \pa_i^{\overline{\pi}}, \pa^{\cal G}_i \setminus A) = p^e(V_i \mid a^{\pi} \cap \pa^{{\cal G}^e}_i, \pa^{{\cal G}^e}_i \setminus A).
\]
Though, as discussed previously, if the graph $\G$ is an NPSEM-IE the graph ${\cal G}^e$ may be regarded as a purely formal construction that aids in the derivation of the identifying formulae. However, if ${\cal G}^e$ is interpreted as a causal model, so that there are well-defined counterfactuals associated with intervening
on each of the vertices in ${A}^{\Ch}$, then ${\cal G}^e$ provides an interventionist interpretation of path-specific counterfactuals;
in fact these interventions allow the identification results above to be checked, in principle, in a hypothetical randomized trial.


In the causal models derived from DAGs with unobserved variables (e.g.,\ ${\cal G}(V \cup H)$), identification of distributions on potential outcomes
such as $p(V(a))$ or $p(V(\pi,a,a'))$ may be stated without loss of generality on the latent projection ADMG ${\cal G}(V)$.  
A complete algorithm for identification of path-specific effects in NPSEM-IEs with hidden variable was given in \citep{shpitser13cogsci} and presented in a more concise form in \citep{shpitser18medid}.  

We now show that identification theory for $p(V(\pi,a,a'))$ in latent projection ADMGs ${\cal G}(V)$ may be restated, without loss of generality, in terms of identification of $p^e(V(a^{\pi}))$ in ${\cal G}^e(V \cup A^{\Ch})$.


\begin{prop} Let ${\cal G}(V \cup H)$ be a DAG. Under the  FFRCISTG model associated with the edge expanded DAG ${\cal G}^e(V \cup A^{\Ch}\cup H)$,
	for any edge consistent $\pi,a,a'$ and $Y \subseteq V$, it follows that $Y(\pi,a,a')=Y(a^{\pi})$ and thus 
$p(Y(\pi,a,a'))$ is identified given $p(V)$ if and only if
$p^e(Y(a^{\pi}))$ is identified from $p^e(V \cup A^{\Ch})$.
\label{prop:e-id-iff}
\end{prop}
This Proposition is a generalization of Theorem \ref{thm:dag-ext-ffrcistg}  from DAGs to latent projection ADMGs.
To determine whether $p^e(Y(a^{\pi}))$ is identified we examine the ADMG ${\cal G}^e(V, A^{\Ch})$. Since this graph is a standard latent projection ADMG (albeit with deterministic relationships relating $A$ and $A^{\Ch}$), the extended ID algorithm decomposes 
the distribution $p^e(Y(a^{\pi}))$  into a set of factors as in (\ref{eqn:line-4}).
However, in order for $p^e(Y(a^{\pi}))$ to be identified 
given data $p(V)$,\footnote{Note that the vertex set for ${\cal G}^e(V, A^{\Ch})$ comprises $V\cup A^{\Ch}$. Further, by construction, the distribution over $p^e(V\cup A^{\Ch})$ is degenerate since the variables in $A^{\Ch}$ are deterministic functions of those in $A$.} an additional requirement must be placed on the terms of this decomposition.  Specifically, for each term $p^e(V_D(a^{\pi},v_{\pas_{D}^{{\cal G}^e(Y(a^{\pi}))}}) = v_{D})$ in (\ref{eqn:line-4}), it must be the case that $a^{\pi}$ assigns consistent values to each element of $A$ that is in $\pas_{D}^{{\cal G}^e(Y(a^{\pi}))}$.  This requirement corresponds to the \emph{recanting district criterion}
which was introduced and shown to be complete in \cite{shpitser13cogsci}.\label{page:recanting}  Aside from this requirement, each term must be identified by the extended ID algorithm described in 
\citep{rrs21volume_id_arxiv}.\footnote{See point (b) in Section \ref{subsec:generalization3}.}


\subsection{Conditional Path-Specific Distributions}

Having established that we can identify path-specific effects by working with potential outcomes derived from the $\mathcal{G}^e$ model, we turn to the identification of conditional path-specific effects using the po-calculus. In \cite{shpitser06idc}, the authors present the conditional identification (IDC) algorithm for identifying quantities of the form $p(Y(x)|W(x))$ (in our notation), given an ADMG. Since conditional path-specific effects correspond to exactly such quantities defined on the 
model associated with the edge expanded graph $\mathcal{G}^e$, we can leverage their scheme for our purposes. The idea is to reduce the conditional problem, identification of $p^e(Y(a^{\pi})|W(a^{\pi}))$, to an unconditional (joint) identification problem for which a complete identification algorithm already exists.

The algorithm has three steps: first, exhaustively apply Rule $2$ of the po-calculus to reduce the conditioning set as much as possible; second, identify the relevant joint distribution using Proposition \ref{prop:e-id-iff} and the complete algorithm in \citep{shpitser18medid}; third, divide that joint by the marginal distribution of the remaining conditioning set to yield the conditional path-specific potential outcome distribution.

Note that we make use of SWIGs defined from edge expanded graphs of the form ${\cal G}^e(a^{\pi},z)$. Beginning with ${\cal G}^e$ the SWIG ${\cal G}^e(a^{\pi},z)$ is constructed by the usual node-splitting operation: split nodes $Z$ and $A_i^j$ into random and fixed halves, where $A_i^j$ has a fixed copy $a$ if $A_i \to V_j$ in ${\cal G}(V)$ is in $\pi$, and $a'_i$ if $A_i \to V_j$ in ${\cal G}(V)$ is not in $\pi$. Relabeling of random nodes proceeds as previously described.
This procedure is in fact complete, as shown by the following result with a proof found in \cite{malinsky19po}.\footnote{\cite{malinsky19po} assumed an NPSEM-IE but the proof only uses the rules of the po-calculus, hence also applies under the less restrictive FFRCISTG assumptions for $\G^e$.}

\begin{thm}
\label{thm:maximal}
	Let $p(Y(\pi,a,a') \,|\, W(\pi,a,a'))$ be a conditional path-specific distribution in the causal model for ${\cal G}$, and let
	$p^e(Y(a^{\pi}) \,|\, W(a^{\pi}))$ be the corresponding distribution under the FFRCISTG associated with the edge expanded graph ${\cal G}^e(V \cup A^{\Ch})$.
	Let $Z$ be the maximal subset of $W$ such that $p^e(Y(a^{\pi}) \,|\, W(a^{\pi})) = p^e(Y(a^{\pi},z) \,|\, W(a^{\pi},z) \setminus Z(a^{\pi},z))$.
	Then $p^e(Y(a^{\pi}) \,|\, W(a^{\pi}))$ is identifiable in ${\cal G}^e$ if and only if $p^e(Y(a^{\pi},z), W(a^{\pi},z) \setminus Z(a^{\pi},z))$ is 
	identifiable in ${\cal G}^e$.
\end{thm}

\begin{figure*}
	\begin{center}
		\begin{tikzpicture}[>=stealth, node distance=1.2cm]
		\tikzstyle{format} = [draw, very thick, circle, minimum size=5.0mm,
		inner sep=0pt]
		\tikzstyle{unode} = [draw, very thick, circle, minimum size=1.0mm,
		inner sep=0pt]
		\tikzstyle{square} = [draw, very thick, rectangle, minimum size=4mm]
		
		\begin{scope}[xshift=0.0cm]
		\path[->, very thick]
		node[] (dummy) {}
		node[format, below of=dummy, yshift=0.4cm] (c) {$C$}
		node[format, right of=c] (w) {$A$}
		node[format, right of=w] (m) {$M$}
		node[format, right of=m] (y) {$Y$}
		
		node[format, gray, above of=w, yshift=-0.5cm] (h1) {$H_1$}
		node[format, gray, below of=w, xshift=0.6cm, yshift=+0.5cm] (h2) {$H_2$}
		
		(c) edge[blue] (w)
		(w) edge[blue] (m)
		(m) edge[blue] (y)
		
		(h1) edge[red] (c)
		(h1) edge[red] (m)
		
		(h2) edge[red] (c)
		(h2) edge[red] (y)
		
		(w) edge[blue, bend left] (y)
		
		node[below of=w, yshift=-0.3cm, xshift=0.6cm] (l) {$(a)$}
		;
		\end{scope}
		
		\begin{scope}[xshift=5.0cm]
		\path[->, very thick]
		node[] (dummy) {}
		node[format, below of=dummy, yshift=0.4cm] (c) {$C$}
		node[format, right of=c] (w) {$A$}
		node[format, right of=w] (m) {$M$}
		node[format, right of=m] (y) {$Y$}
		
		node[format, gray, below of=w, xshift=0.6cm, yshift=+0.5cm] (h2) {$H_2$}
		
		(c) edge[blue] (w)
		(w) edge[blue] (m)
		(m) edge[blue] (y)
		
		
		(c) edge[blue, bend left] (m)
		
		(h2) edge[red] (c)
		(h2) edge[red] (y)
		
		(w) edge[blue, bend left] (y)

		node[below of=w, yshift=-0.3cm, xshift=0.6cm] (l) {$(b)$}
		;
		\end{scope}
		
		\end{tikzpicture}
	\end{center}
	\caption{(a) A hidden variable causal DAG where $p(Y(a,M(a')))$ is identified, but $p(Y(a,M(a')) \mid C)$ is not identified.
		(b) A seemingly similar hidden variable causal DAG where both $p(Y(a,M(a')))$ and $p(Y(a,M(a')) \mid C)$ are identified.
	}
	\label{fig:ex-po-calc}
\end{figure*}
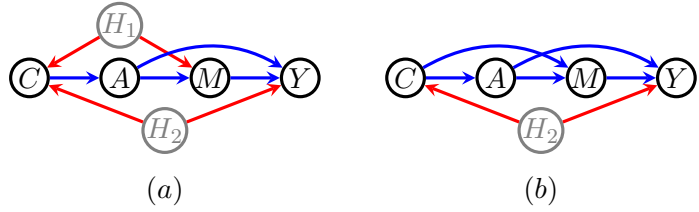

As an example, $p(Y(a,M(a')))$ is identified from $p(C,A,M,Y)$ in the causal model in Figure~\ref{fig:ex-po-calc}~(a), via
\[
p(Y(a,M(a')))=\sum_{m} \left(\frac{
\sum_{c} p(Y,m \mid a, c) p(c) 
}{
\sum_{c} p(m \mid a, c) p(c)
}\right)\left(\sum_{c^*} p(m \mid a', c^*) p(c^*)\right).
\]
However $p(Y(a,M(a')) | C)$ is not identified, since 
$p(Y(a,M(a')), C)$ must first be identified, and this joint distribution is not identified via results in \cite{shpitser13cogsci}. 
On the other hand, $p(Y(a,M(a')) | C)$ is identified from $p(C,A,M,Y)$ in a seemingly similar graph in Figure~\ref{fig:ex-po-calc}~(b), via
$\sum_{M} p(Y\mid M,a, C) p(M \mid a',C)$.

\section{Conclusion}
We have shown here, that graphical insights, derived from Pearl's thought experiment, 
may be used to place mediation analysis on a firm interventionist footing, and yield analyses of direct, indirect, and path-specific effects that are amenable to falsification, and explainable to practitioners.
%


{

\begingroup
\let\cleardoublepage\clearpage
\bibliography{references}
\endgroup
}

\newpage

\section{Appendix}

\subsection{Proof Of PDE Bounds Under The FFRCISTG Model}

We here prove the bounds on the PDE implied by the FFRCISTG model associated with the graph in Figure~\ref{fig:amyno}~(a).

\begin{prf} 
It follows from the definition that:
$PDE_{a,a'}=P( Y(a,M(a'))=1) -P(Y=1\mid A=a').$
Note  that 
\begin{eqnarray*}
p(Y(a,M(a'))\!=\!1) &=&p(Y(a,m=0)\!=\!1\mid M(a')=0) p(M(a')=0\mid A=a') \\
&&\kern10pt+p(Y(a,m=1)\!=\!1 \mid M(a')=1) p(M(a')=1\mid A=a')\\
&=&p(Y(a,m=0)\!=\!1\mid M(a')=0) p(M=0\mid A=a') \\
&&\kern10pt+p(Y(a,m=1)\!=\!1 \mid M(a')=1) p(M=1\mid A=a').
\end{eqnarray*}%
The quantities $p(Y(a,m=0)\!=\!1\,|\, M(a')=0)$ and $p(Y(a,m=1)\!=\!1\,|\,
M(a')=1)$ are constrained by the law for the observed data via: 
\begin{eqnarray*}
p(Y\!=\!1\,|\,A\!=\!a,M\!=\!0) &=&p(Y(a,m=0)\!=\!1) \\
&=&p(Y(a,m=0)\!=\!1\mid M(a')=0)p(M(a')=0) \\
&&\kern10pt+p(Y(a,m=0)\!=\!1\mid M(a')=1)p(M(a')=1) \\
&=&p(Y(a,m=0)\!=\!1\mid M(a')=0)p(M=0\mid A=a') \\
&&\kern10pt+p(Y(a,m=0)\!=\!1\mid M(a')=1)p(M=1\mid A=a'),
\end{eqnarray*}%
\begin{eqnarray*}
p(Y\!=\!1\,|\,A\!=\!a,M\!=\!1) &=&p(Y(a,m=1)\!=\!1) \\
&=&p(Y(a,m=1)\!=\!1\mid M(a')=0)p(M(a')=0) \\
&&\kern10pt+p(Y(a=1,m=1)\!=\!1\mid M(a')=1)p(M(a')=1) \\
&=&p(Y(a,m=1)\!=\!1\mid M(a')=0)p(M=0\mid A=a') \\
&&\kern10pt+p(Y(a,m=1)\!=\!1\mid M(a')=1)p(M=1\mid A=a').
\end{eqnarray*}

It then follows from the analysis in Section 2.2 in \cite{richardson:2010} that
the set of possible values for the pair 
\begin{equation*}
(\alpha_0,\alpha_1) \equiv \left(p( Y( a,m\!=\!0 )=1 \,|\, M(a') =0 ), p(Y(
a,m\!=\!1 )=1 \,|\, M(a') =1 )\right) 
\end{equation*}
compatible with the observed joint distribution $p(m,y \mid a)$ is given by: 
\begin{equation*}
 (\alpha_0, \alpha_1) \;\in\; [l_0, u_0] \times [l_1, u_1] 
\end{equation*}
where, 
\begin{eqnarray*}
l_0 &=& \max\{0, 1+ (p(Y\!=\!1\mid A=a,M=0) -1)/p(M=0 \mid A=a')\}, \\
u_0 &=& \min\{ p(Y\!=\!1\mid A=a,M=0)/p(M=0 \mid A=a'),1\}, \\[12pt]
l_1 &=& \max\{0, 1+ (p(Y\!=\!1\mid A=a,M=1) -1)/p(M=1\mid A=a')\}, \\
u_1 &=& \min\{p(Y\!=\!1\mid A=a,M=1)/p(M=1 \mid A=a'),1\}.\\[-36pt]
\end{eqnarray*}
\end{prf}

\subsection{Proof That The PDE Is Not Identified In The River Blindness Study}\label{subsec:pde-not-identified-river-blindness}

\begin{figure}
	\begin{center}
		\begin{tikzpicture}[>=stealth, node distance=1.2cm]
		\tikzstyle{format} = [draw, thick, circle, minimum size=4.0mm,
		inner sep=1pt]
		\tikzstyle{unode} = [draw, thick, circle, minimum size=1.0mm,
		inner sep=0pt,outer sep=0.9pt]
		\tikzstyle{square} = [draw, very thick, rectangle, minimum size=4mm]

	\begin{scope}[xshift=0cm]
		\path[->,  line width=0.9pt]
		node[format, shape=ellipse] (a) {$A$}
		node[format, shape=ellipse, above right of=a, yshift=-0.3cm, fill=lightgray] (s) {$S$}			
		node[format, shape=ellipse, below right of=a, yshift=+0.3cm] (m) {$M$}
		
		node[format, shape=ellipse, left= of s, xshift=0.25cm, yshift=0.5cm, fill=lightgray] (u) {$U$}

		node[format, shape=ellipse, right of=m,xshift=-0.0cm] (y) {$Y$}

		(a) edge[blue] (s)
		(a) edge[blue] (m)
		(s) edge[blue] (y)
		(m) edge[blue] (y)
		
		(u) edge[blue, out=270,in=180] (m)
		(u) edge[blue, out=0,in=90] (y)

		;
	\end{scope}
		\end{tikzpicture}
		\end{center}
\caption{
The DAG corresponding to the projection of the DAG $\cal G$ shown in Figure~\ref{fig:torpedo3}~(a) after marginalizing $R$.
\label{fig:pdewashup}}
\end{figure}
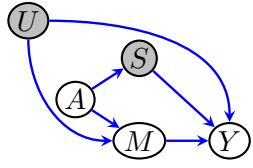

Consider the NPSEM-IE corresponding to the DAG shown in Figure~\ref{fig:pdewashup}, which may be obtained by marginalizing $R$ from the causal
DAG representing the River Blindness studies shown in Figure~\ref{fig:torpedo3}(a).

The one-step-ahead counterfactuals defining the model are: $U$, $A$, $S(a)$, $M(a,u)$, $Y(m,s,u)$.
We will assume that all variables, including $U$, are binary. Though not shown explicitly on the graph, we also have the following constraints: (i) $Y(m,s_0,u_0)= Y(m,s_0,u_1)\equiv Y(m,s_0)$;
(ii) $Y(m_0,s_1,u_0) = Y(m_0,s_1,u_1)\equiv Y(m_0,s_1)$;
(iii)
$M(a_1,u_0)=M(a_1,u_1)\equiv M(a_1)$, and (iv) $S(a)=a$, where here we use $x_i$ as a shorthand for $x=i$. These arise from, respectively, (i) the fact that if immuno-suppressants are not available ($s=0$), then the patient's outcome ($Y$) is not influenced by their predisposition  ($U$) to use medicine; (ii) the availability of immuno-suppressants, and hence the patient's predisposition ($U$) to use them, is not relevant to patients who do not receive Ivermectin;\footnote{This is an additional restriction not present in the original account, but introduced here solely to reduce the number of parameters. This corresponds to removing the $U\rightarrow R$ edge in the SWIG ${\cal G}(a_1,m_0,s_1)$ where $m$ is set to $0$; see Figure~\protect\ref{fig:g-of-asm}.
(Since non-identifiability under a submodel implies non-identifiability in the larger model, we can make this assumption without loss of generality.)}
 (iii) the fact that in a randomized trial ($a=1$) the treatment the patient receives is not influenced by $U$; (iv) The fact that the clinic is available  ($s=1$) if and only if a patient is in the randomized trial ($a=1$). The NPSEM-IE is then defined by the following $10$ parameters:
\begin{align*}
U: && \theta_U &\equiv p(U=1);\\
A: && \theta_A &\equiv p(A=1);\\[4pt]
M(a,u): && \theta_M(a_0,u) &\equiv p(M(a_0,u) = 1),&&\hbox{ for }u\in\{0,1\},\\
&&  \theta_M(a_1) &\equiv p(M(a_1) = 1);\\[4pt]
Y(m,s,u): && \theta_Y(m_1,s_1,u) &\equiv p(Y(m_1,s_1,u) = 1),&&\hbox{ for }u\in\{0,1\},\\
&&  \theta_Y(m,s) &\equiv   p(Y(m,s) = 1),&&\hbox{ for }(m,s)\in\{(0,0),(0,1),(1,0)\}.
\end{align*}
Let ${\mathbf \theta}$ denote a vector containing all of these parameters.
The corresponding observed distribution $p_{\bf\theta}(A,M,Y)$ is given by the following equations:
\begin{align*}
p_{\bf\theta}(A=1) &= \theta_A,\\[4pt]
p_{\bf\theta}(M=1\,|\,A=a)&= \left\{\begin{array}{cr}
 \theta_M(a_0,u_0)(1-\theta_U) +  \theta_M(a_0,u_1)\theta_U,& \hbox{ if } a=0,\\
  \theta_M(a_1), &\hbox{ if } a=1,
  \end{array}
  \right.
  \end{align*}
  \vspace{-15pt}
  \begin{align*}
 \MoveEqLeft[4]{p_{\bf\theta}(Y=1\,|\,A=a, M=m)}\\
 & = \left\{\begin{array}{cl}
 \theta_Y(m,s_a),& \hbox{if }(a,m)\in\{(0,0),(0,1),(1,0)\},\\
  \theta_Y(m_1,s_1,u_0)(1-\theta_U) + \theta_Y(m_1,s_1,u_1)\theta_U,  &\hbox{ if } (a,m)=(1,1). 
  \end{array}
  \right.
\end{align*}
Given $\theta$, consider a perturbed vector $\tilde{\theta}$ defined by taking
\begin{align*}
\tilde{\theta}_M(a_0,u_0) &\equiv \theta_M(a_0,u_0) + \epsilon/(1-\theta_U),\\
\tilde{\theta}_M(a_0,u_1) &\equiv \theta_M(a_0,u_1) - \epsilon/\theta_U,
\end{align*}
for sufficiently small $\epsilon$ and leaving the other $8$ parameters unchanged from $\theta$.
It is simple to see that the resulting observed distribution  is unchanged, so  $p_{{\bf\theta}_\epsilon}(A,M,Y) = p_{\bf\theta}(A,M,Y)$, since the perturbation only changes the expression for
$p(M=1\,|\,A=0)$, but the additional terms involving $\epsilon$ cancel.

Turning to the PDE we see that:
\begin{align*}
\MoveEqLeft{p_{\theta}(Y(a_1,M(a_0)) = 1)}\\
&= p_{\theta}(Y(M(a_0),s_1)=1)\\
&= \sum_{u} p_{\theta}(Y(M(a_0),s_1)=1\,|\, U=u)p_{\theta}(U=u)\\
&= \sum_{u,k} p_{\theta}(Y(m_k,s_1)=1\,|\, U=u, M(a_0)=k)p_{\theta}(M(a_0) = k\,|\, U=u)p_{\theta}(U=u)\\
\hbox{\tiny consistency}&= \sum_{u,k} p_{\theta}(Y(m_k,s_1,u)=1\,|\, U=u, M(a_0)=k)p_{\theta}(M(a_0,u) = k\,|\, U=u)p_{\theta}(U=u)\\
\hbox{\tiny independence}&= \sum_{u,k} p_{\theta}(Y(m_k,s_1,u)=1)p_{\theta}(M(a_0,u) = k)p_{\theta}(U=u)\\
&=\theta_{Y}(m_0,s_1)(1-\theta_M(a_0,u_0))(1-\theta_U) + \theta_{Y}(m_0,s_1)(1-\theta_M(a_0,u_1))\theta_U\\
&\quad+\theta_{Y}(m_1,s_1,u_0)\theta_M(a_0,u_0)(1-\theta_U) + \theta_{Y}(m_1,s_1,u_1)\theta_M(a_0,u_1)\theta_U.
\end{align*}
A simple calculation shows that
\[
p_{\tilde{\theta}}(Y(M(a_0),s_1)=1) = p_{\theta}(Y(M(a_0),s_1)=1) + \epsilon\left(\theta_{Y}(m_1,s_1,u_0) - \theta_{Y}(m_1,s_1,u_1)\right)
\]
so that the PDE will take a different value under $p_{\tilde{\theta}}$, so long as $\theta_{Y}(m_1,s_1,u_0) \neq \theta_{Y}(m_1,s_1,u_1)$.\footnote{This inequality corresponds to the presence of the edge $U\rightarrow Y$ in Figure~\ref{fig:pdewashup}.}

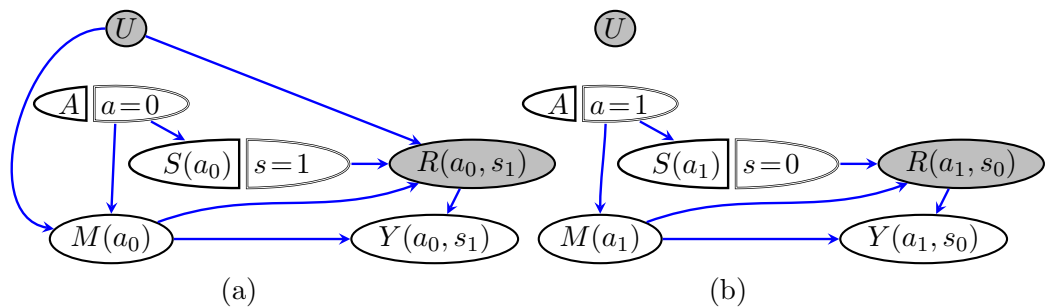
\begin{figure}
	\begin{center}
		\begin{tikzpicture}[>=stealth, node distance=1.2cm]
		\tikzstyle{format} = [draw, thick, circle, minimum size=4.0mm,
		inner sep=1pt]
		\tikzstyle{unode} = [draw, thick, circle, minimum size=1.0mm,
		inner sep=0pt,outer sep=0.9pt]
		\tikzstyle{square} = [draw, very thick, rectangle, minimum size=4mm]
	\begin{scope}
		\begin{scope}
			\tikzset{line width=0.9pt, inner sep=2pt, swig vsplit={gap=3pt, inner line width right=0.3pt}}	
				\node[ xshift=0.0cm, yshift=0.0cm, name=a, shape=swig vsplit]{
        					\nodepart{left}{$A$}
        					\nodepart{right}{$a\!=\!0\,$} };
		\end{scope}
		\begin{scope}
			\tikzset{line width=0.9pt, inner sep=2pt, swig vsplit={gap=3pt, inner line width right=0.3pt}}	
				\node[right of=a, xshift=0.5cm, yshift=-0.8cm, name=s, shape=swig vsplit]{
        					\nodepart{left}{$S(a_0)$}
        					\nodepart{right}{$s\!=\!1\,$} };
		\end{scope}
		\path[->,  line width=0.9pt]
		node[format, shape=ellipse, below of=a, xshift=0cm, yshift=-0.6cm] (m) {$M(a_0)$}
		node[format, shape=ellipse, above of=  a, xshift=0.2cm, yshift=-0.2cm, fill=lightgray] (u) {$U$}
		node[format, shape=ellipse, right of=s, xshift=1.9cm, fill=lightgray] (r) {$R(a_0,s_1)$}
		node[format, shape=ellipse, right of=m, xshift=3.1cm] (y) {$Y(a_0,s_1)$}
		(a) edge[blue] (s)
		(a) edge[blue] (m)
		(s) edge[blue] (r)
		(m) edge[blue] (y)
		(r) edge[blue] (y)
		(u) edge[blue, out=180,in=170] (m)
		(u) edge[blue] (r)
		(m) edge[blue, out=20,in=200] (r)
		node[below of=s,yshift=-0.5cm]{(a)}	
		;
		\end{scope}
	\begin{scope}[xshift=6.5cm]
		\begin{scope}
			\tikzset{line width=0.9pt, inner sep=2pt, swig vsplit={gap=3pt, inner line width right=0.3pt}}	
				\node[ xshift=0.0cm, yshift=0.0cm, name=a, shape=swig vsplit]{
        					\nodepart{left}{$A$}
        					\nodepart{right}{$a\!=\!1\,$} };
		\end{scope}

		\begin{scope}
			\tikzset{line width=0.9pt, inner sep=2pt, swig vsplit={gap=3pt, inner line width right=0.3pt}}	
				\node[right of=a, xshift=0.5cm, yshift=-0.8cm, name=s, shape=swig vsplit]{
        					\nodepart{left}{$S(a_1)$}
        					\nodepart{right}{$s\!=\!0\,$} };
		\end{scope}

		\path[->,  line width=0.9pt]
		node[format, shape=ellipse, below of=a, xshift=0cm, yshift=-0.6cm] (m) {$M(a_1)$}
		node[format, shape=ellipse, above of= a, xshift=0.2cm, yshift=-0.2cm, fill=lightgray] (u) {$U$}

		node[format, shape=ellipse, right of=s, xshift=1.9cm, fill=lightgray] (r) {$R(a_1,s_0)$}
		node[format, shape=ellipse, right of=m, xshift=3.1cm] (y) {$Y(a_1,s_0)$}

		(a) edge[blue] (s)
		(a) edge[blue,out=270,in=90] (m)
		(s) edge[blue] (r)
		(m) edge[blue] (y)
		(r) edge[blue] (y)
		
		(m) edge[blue, out=20,in=200] (r)

		node[below of=s,yshift=-0.5cm]{(b)}	
		;
		\end{scope}
		\end{tikzpicture}
		\end{center}	
\caption{(a) The SWIG $\G(a=0,s=1)$ associated with the DAG shown in Figure~\protect\ref{fig:torpedo3}~(a);
(b) The SWIG $\G(a=1,s=0)$ associated with the DAG shown in Figure~\protect\ref{fig:torpedo3}~(a).
The $U\rightarrow M$ edge is absent because this is a randomized study ($a=1$).
The $U\rightarrow R$ edge is not present because the clinic is not available ($s=0$), hence patients do not have the option
to take immuno suppressants.
\label{fig:g-of-asm2}}
\end{figure}

That the PDE is not identified in the River Blindness study example should not be surprising in light of the results in Section \ref{sec:path-specific}.
In particular, we know that $Y(a_1, M(a_0)) = Y(s_1,a_0)$.\footnote{This assumes that missing edges correspond to the absence of direct effects at the individual level,
so that $M(a_0,s_1) = M(a_0)$.} In addition, we see that in the SWIG $\G(a\!=\!0,s\!=\!1)$ shown in Figure~\ref{fig:g-of-asm2}~(a),  which here plays the role of the SWIG for the expanded graph ${\cal G}^{ex}$, $Y(a_0,s_1)$ is in the same district as $M(a_0)$, but the fixed nodes $a=0$ and $s=1$ are both parents of this district, but are both associated with $A$ in the original graph. Consequently $M(a_0)\leftrightarrow Y(a_0,s_1)$ forms a recanting district.
However, formally Section~\ref{sec:path-specific} considers models defined solely via ordinary conditional independences, whereas the model in the river blindness study also incorporated context specific independences. For this reason, since a quantity may be unidentified in a model, but identified in a submodel, we provided an explicit construction of an NPSEM-IE for the DAG in Figure~\ref{fig:torpedo3}~(a).\footnote{Since the NPSEM-IE is a submodel of the FFRCISTG, this also establishes that the PDE is not identified under the latter interpretation of the DAG in Figure~\ref{fig:torpedo3}~(a).}

Similarly, consideration of the SWIG $\G(a\!=\!1,s\!=\!0)$ shown in Figure~\ref{fig:g-of-asm2}~(b), shows that $p(Y(a_1,s_0))$ is identified from $p(A,M,Y)$ because there is no recanting district:\footnote{Note that this identification argument does not use the additional constraint (ii).}
\begin{align}
\notag
p(Y(a_1,s_0) = 1) &= \sum_{r,m} p(y=1\,|\, r,m) p(r\,|\, s=0,m)p(m\,|\,a=1)\quad \hbox{\tiny g-formula}\\
\notag
\hbox{\tiny independence} &= \sum_{r,m} p(y=1\,|\, r,m,s=0) p(r\,|\, s=0,m)p(m\,|\,a=1)\\
\notag
\hbox{\tiny marginalization} &= \sum_{m} p(y=1\,|\, m,s=0)p(m\,|\,a=1)\\
\label{eqn:determinism}
\hbox{\tiny determinism}&= \sum_{m} p(y=1\,|\, m,a=0)p(m\,|\,a=1).
\end{align}
Thus, in this example, $p(Y(a,s))= p(Y(s,M(a)))$ is identified for $(a,s) \in \{(0,0),(1,1),(1,0)\}$, but is not identified for $(a,s) = (0,1)$.
Since $Y(a_0,M(a_1)) =Y(a_1,s_0)$,  it follows that $p(Y(a_0,M(a_1)))$, which forms part of the Total Direct Effect (\ref{eq:tde-intro}),  is also identified.

This conclusion also follows from Proposition \ref{prop:2-to-4}, here letting $A$ be ``$N$'' and $S$ be ``$O$,'' since equality (\ref{eq:conditions-for-4-to-2a}) holds with $x=1$ (though not with $x=0$).

\subsection{Detecting Confounding via Interventions on $A$ and $S$}\label{subsec:detectable-confounding}

In the river blindness study, it is not possible to detect the confounder $U$ via interventions on $A$ and $M$, since the distribution $\{A,M(a),Y(a,m)\}$ obeys the 
FFRCISTG model corresponding to Figure~\ref{fig:amyno}~(a). 
This is due to the context-specific independences that hold in this example;  see Section \ref{subsec:ffrcistg-not-npsem-ie}.
However, confounding becomes detectable if we are able to intervene on $S$ and $A$. To see this note that:

\begin{align*}
p(Y(a_0,s_1)=1 \,|\, M(a_0) =1]&= \sum_{u}p(Y(a_0,s_1)=1 \,|\, M(a_0) =1,U=u)p(u\,|\,M(a_0)=1) \\
&= \sum_{u}p(Y(m_1,s_1,u)=1 \,|\, M(a_0) =1,U=u)p(u\,|\,M(a_0)=1) \\
&=\sum_{u} p(Y(m_1,s_1,u)=1) p(u\,|\,M(a_0)=1)\\
&=\frac{\theta_{Y}(m_1,s_1,u_0)(1-\theta_U)\theta_M(a_0,u_0) + \theta_{Y}(m_1,s_1,u_1)\theta_U\theta_M(a_0,u_1)}{(1-\theta_U)\theta_M(a_0,u_0) + \theta_U\theta_M(a_0,u_1)}
\end{align*}
which will depend on $\theta_M(a_0,u_0)$ and  $\theta_M(a_0,u_1)$,
(if $\theta_{Y}(m_1,s_1,u_0)\neq \theta_{Y}(m_1,s_1,u_1)$).

However,  
\begin{align*}
p(Y(a_1,s_1) \,|\, M(a_1) =1)&= \sum_{u}p(Y(a_1,s_1) \,|\, M(a_1) =1,U=u)p(u\,|\,M(a_1)=1) \\
&= \sum_{u}p(Y(m_1,s_1,u) \,|\, M(a_1) =1,U=u)p(u\,|\,M(a_1)=1) \\
&=\sum_{u} p(Y(m_1,s_1,u)) p(u\,|\,M(a_1)=1)\\
&=\sum_{u} p(Y(m_1,s_1,u)) p(u)\\
&=\theta_{Y}(m_1,s_1,u_0)(1-\theta_U) + \theta_{Y}(m_1,s_1,u_1)\theta_U.
\end{align*}
Thus $p(Y(a_0,s_1)=1 \,|\, M(a_0) =1) \neq p(Y(a_1,s_1)=1 \,|\, M(a_1) =1)$, while if $U$ were absent we would have equality.
\footnote{Note that under the additional assumption (iii), we have: $p(Y(a_0,s_1)=1 \,|\, M(a_0) =0) = p(Y(a_1,s_1)=1 \,|\, M(a_1) =0)$
since $\theta_{Y}(m_0,s_1,u_0)= \theta_{Y}(m_0,s_1,u_1)$. Without (iii) the equality will not hold.}
That the equality holds if $U$ is absent may be seen by removing $U$ from the SWIG ${\cal G}(a,s)$ constructed from $\cal G$ in Figure~\ref{fig:torpedo3}~(a)  and then applying Rule 3 of the po-calculus \citep{rrs21volume_id_arxiv}: the equality follows from the fact that $a$ is d-separated from $Y(a,s)$ given $M(a)$.
Conversely, that the equality fails to hold when $U$ is present is not surprising since 
we have the d-connecting path $a \rightarrow M(a) \leftarrow U \rightarrow R(m,s) \rightarrow Y(m,s)$ in  Figure~\ref{fig:g-of-asm2}~(a).

\subsection{Proof of Proposition \ref{prop:equiv}}

\begin{prf}
For any $V_i \in V$, $V_i(a^{\pi})$ is defined via (\ref{eqn:rec-sub}) applied to ${\cal G}^{{edge}}$,
\begin{align*}
V_i(a^{\pi}) \equiv V_i\!\left(a^{\pi}_{\pa^{{\cal G}^{{edge}}}_i \cap A^{\Ch}},\, V_{ \pa^{{\cal G}^{{edge}}}_i\! \setminus A^{\Ch}}(a^{\pi})\right).
\end{align*}
Similarly, $V_i(\pi,a,a')$ is defined via (\ref{eqn:pse}) applied to ${\cal G}$ as
\begin{align*}
V_i(\pi, a, a') &\equiv a \text{ if }V_i \in A,\\
V_i(\pi, a, a') &\equiv 
V_i( \{ V_j(\pi, a, a') \mid V_j \in \pa^{\pi}_i \}, \{ V_j(a') \mid V_j \in \pa^{\overline{\pi}}_i \} )
\end{align*}
where $V_j(a') \equiv a'$ if $V_j \in A$, and given by (\ref{eqn:rec-sub}) otherwise, 
$\pa^{\pi}_i$ is the set of parents of $V_i$ along an edge which is a part of a path in $\pi$, and
$\pa^{\overline{\pi}}_i$ is the set of all other parents of $V_i$.

By definition of ${\cal G}^{{edge}}$, the induction on the tree structure of (\ref{eqn:pse}) in ${\cal G}$ matches the induction on the tree structure
of (\ref{eqn:rec-sub}) in ${\cal G}^{{edge}}$.  

Finally, since $V_i(\pi,a,a')$ is edge consistent, any edge of the form $A_k \to V_j$ in ${\cal G}$, for any $A_k \in A$ that starts a proper causal path that ends at $V_i$ is assigned precisely one
value (either $a_{A_k}$ or $a'_{A_k}$).  Moreover, this value in the base case of the induction of (\ref{eqn:pse}), by definition of $a^{\pi}$ matches the value $a^{\pi}_{A_k}$ assigned by the corresponding base case of the induction of (\ref{eqn:rec-sub}).  This establishes our conclusion.

That the corresponding observed data variables $V_i$ match in models represented by the original graph ${\cal G}$, and the edge expanded graph ${\cal G}^{{edge}}$ follows by the above argument, using the fact that all elements in $A^{\Ch}$ are deterministically related to the appropriate elements in $A$.

\end{prf}

 \cite{malinsky19po} prove a weaker result establishing equality in distribution.



\makeatletter\@input{xx_med.tex}\makeatother
\end{document}